\documentclass[aps,reprint]{revtex4-1}
\usepackage{amsmath}
\usepackage{xr-hyper}
\usepackage{hyperref}
\usepackage{isotope}
\usepackage{mhchem}
\usepackage{xcolor}
\usepackage{graphicx}
\usepackage{mathrsfs}
\usepackage{dcolumn}
\usepackage{bm}
\usepackage{cases}
\usepackage{multirow,booktabs}
\usepackage{makecell}
\usepackage{amsmath}
\usepackage{amssymb}
\usepackage{sidecap}
\usepackage{graphicx}
\usepackage{here}
\usepackage{adjustbox}
\usepackage{tikz}
\usepackage{float}
\usepackage{tabularx}
\usepackage{soul}
\usepackage[normalem]{ulem}

\usepackage{xcolor}
\definecolor{pastelgray}{rgb}{0.81, 0.81, 0.77}
\definecolor{beaublue}{rgb}{0.9, 0.9, 0.93}

\newcommand{\nuc}[2]{
    \ensuremath{ {}^{#1} \mathrm{#2} }}

\usepackage{todonotes}

\begin{document}

\title{Deconstructing experimental decay energy spectra: the $^{26}$O case}
\author{Pierre Nzabahimana$^{1}$, Thomas Redpath$^{1,3}$, Thomas Baumann$^1$, Pawel Danielewicz$^1$, Pablo Giuliani$^{1,2}$ and Paul Gu\`eye$^1$}
 \affiliation{$^1$Facility of Rare Isotope Beams and Department of Physics and Astronomy, \\
 	Michigan State University, East Lansing, Michigan 48824, USA}
 \affiliation{$^2$Department of Statistics and Probability, 
 	Michigan State University, East Lansing, Michigan 48824, USA}
 \affiliation{$^3$Virginia State University, Virginia 23806, USA}
\date{\today}


\begin{abstract}

In nuclear reaction experiments, the measured decay energy spectra can give insights into the shell structure of decaying systems. However, extracting the underlying physics from the measurements is challenging due to detector resolution and acceptance effects. The Richardson-Lucy (RL) algorithm, a deblurring method that is commonly used in optics and has proven to be a successful technique for restoring images, was applied to our experimental nuclear physics data. The only inputs to the method are the observed energy spectrum and the detector's response matrix also known as the transfer matrix. We demonstrate that the technique can help access information about the shell structure of particle-unbound systems from the measured decay energy spectrum that isn't immediately accessible via traditional approaches such as chi-square fitting. For a similar purpose, we developed a machine learning model that uses a deep neural network (DNN) classifier to identify resonance states from the measured decay energy spectrum. We tested the performance of both methods on simulated data and experimental measurements. Then, we applied both algorithms to the decay energy spectrum of $^{26}\mathrm{O} \rightarrow ^{24}\mathrm{O}$ + n + n measured via invariant mass spectroscopy. The resonance states restored using the RL algorithm to deblur the measured decay energy spectrum agree with those found by the DNN classifier. Both deblurring and DNN approaches suggest that the raw decay energy spectrum of $^{26}\mathrm{O}$ exhibits three peaks at approximately  0.15~MeV, 1.50~MeV, and 5.00~MeV, with half-widths of 0.29~MeV, 0.80~MeV, and 1.85~MeV, respectively.
\end{abstract}
\pagenumbering{arabic}
\maketitle
\section{Introduction }
Invariant mass spectroscopy allows experimental access to unbound states. However, interpreting and extracting physics from the measured decay energy spectra are often challenged by limited resolution and distortions caused by experimental acceptance effects. This is particularly true in investigations of neutron-unbound states, since they involve the measurement of neutrons and charged decay fragments in coincidence. 
In a decay experiment of this type, the neutron-unbound state is populated through a nuclear reaction induced by a rare isotope beam, typically proton-removal. The unbound state decays immediately, and by measuring the momentum vectors of the decay products, the invariant mass of the unbound system can be calculated. The measured decay energy spectrum can then be reconstructed by subtracting the masses of all constituents of the system. 

\begin{figure*}
    \centering
    \includegraphics[width=0.8\linewidth]{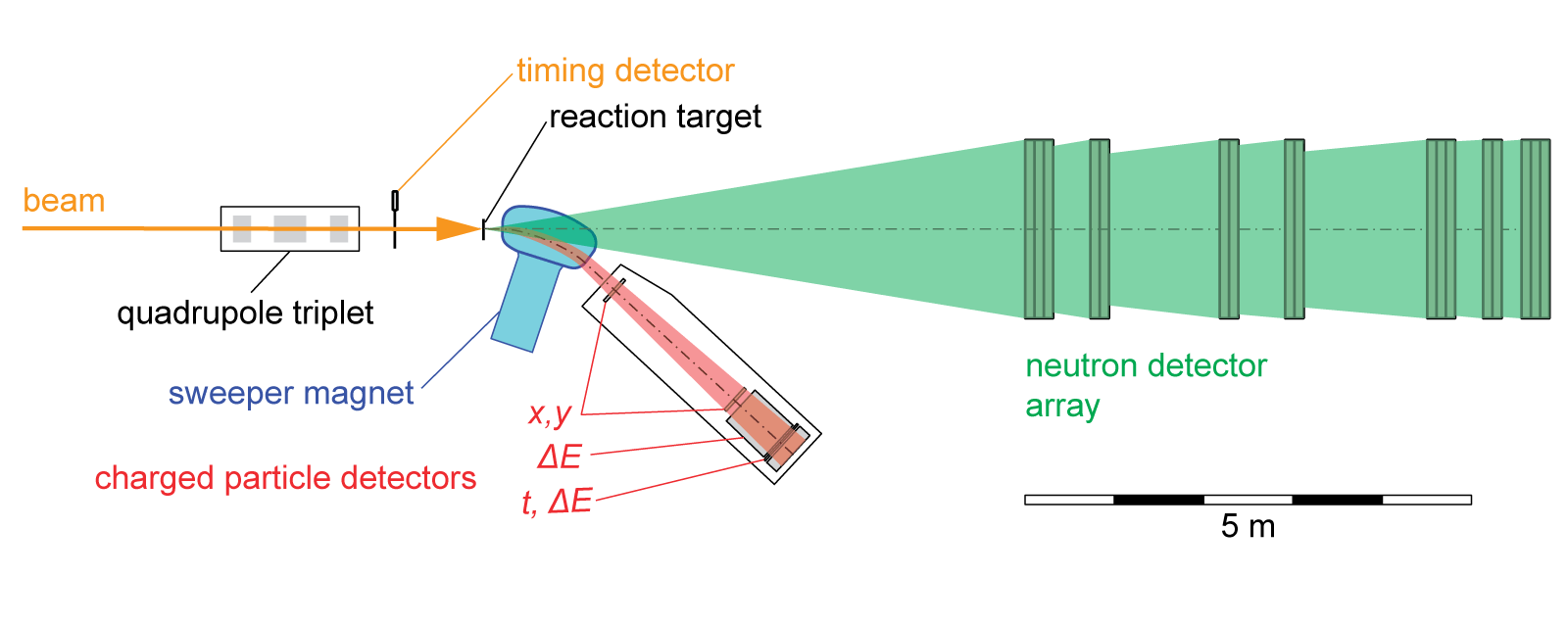}
        \caption{(Color online) The MoNA experimental setup for invariant mass measurements in search of neutron-unbound states includes the Sweeper magnet, charged particle detector suite, and neutron detector array. The rare isotope beam (orange arrow) impinges on a reaction target where the unbound state is populated in a nuclear reaction. The charged breakup fragments (red shaded area) are directed by a magnetic dipole field into the charged particle detector suite, while the neutrons (green shaded area) travel along the beam direction to the neutron detector array.}
    \label{fig:sweeperconcept}
\end{figure*}

In this work, we are focusing on the two-neutron emission decay energy spectrum of $^{26}$O. This unbound nucleus was recently measured by the MoNA Collaboration \cite{redpath2019measuring}, with the setup illustrated in Fig.~\ref{fig:sweeperconcept}. The exploration typifies efforts to learn about the structure of nuclei towards the neutron-drip line~\cite{caesar2013beyond}. In general, measuring neutron momenta implies the use of a neutron detector array that usually has limited position resolution and detection efficiency. Similarly, measuring the momenta of charged particles involves tracking the particle trajectories back through a magnetic field and part of the reaction target to determine the angle and energy at the point of the breakup reaction, which is not accessible to direct measurements. The procedures introduce variations and uncertainties in such a way that the measured decay energy distribution is only a distorted and blurred image of the true decay energy spectrum of the unbound system. In the present work, we will utilize two novel methods of inferring features of the true decay spectrum: a deblurring algorithm and a deep neural network approach. There is much potential for these strategies outside of the particular problem.

The rest of the manuscript is organized as follows.  In Sec.~II, we discuss the interplay of experiment and analysis methods of decay spectrum.  In Sec.~III, we discuss practicalities of deblurring and, in particular, how the procedure is expanded to deal with noisy data. The methodology is tested on simulated data in Sec.~IV. In Sec.~V, we discuss the deblurring of the measured $^{26}$O decay energy spectrum. In Sec.~{VI}, we build a DNN classification model to identify resonance states in the measured decay energy spectrum of $^{26}$O. We present our conclusions and outlook in Sec.~VII.

\section{Interplay of the Experiment and the Analysis Methods}
\subsection{Experiment and Construction of Transfer Matrix}

\begin{figure}[!htb]
	\centering
	\centering
		\includegraphics[width=1\linewidth]{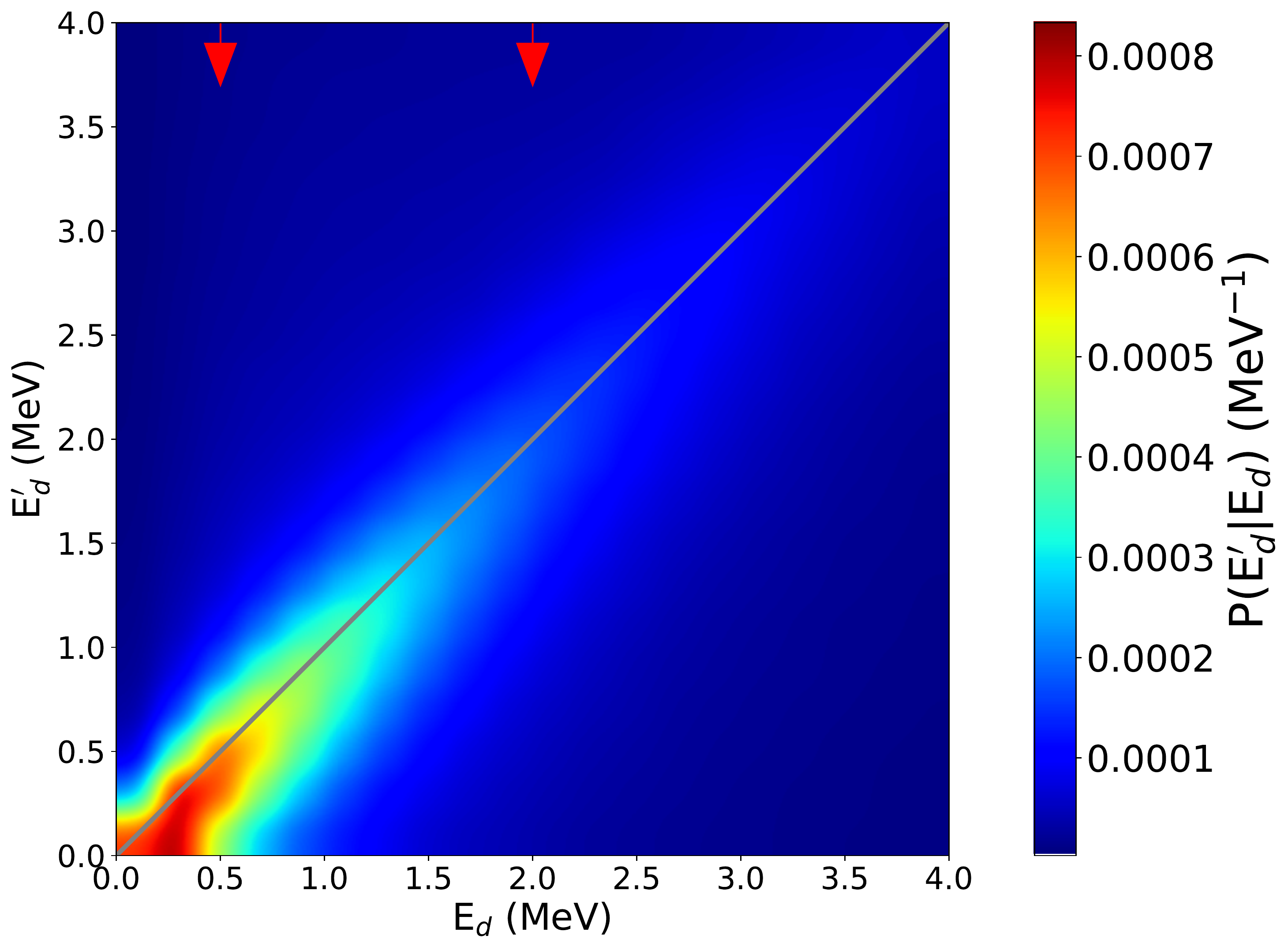}
		\caption{(Color online) The response matrix $P(E_d'|E_d)$ of the MoNA experimental setup depicted in Fig~\ref{fig:sweeperconcept} used in measuring the decay energy of the three-particle decay $^{26}\text{O} \rightarrow ^{24}\text{O} + 2n$. See text for details.}
		\label{fig:TM}
\end{figure}

\begin{figure}
    \centering
		\includegraphics[width=1\linewidth]{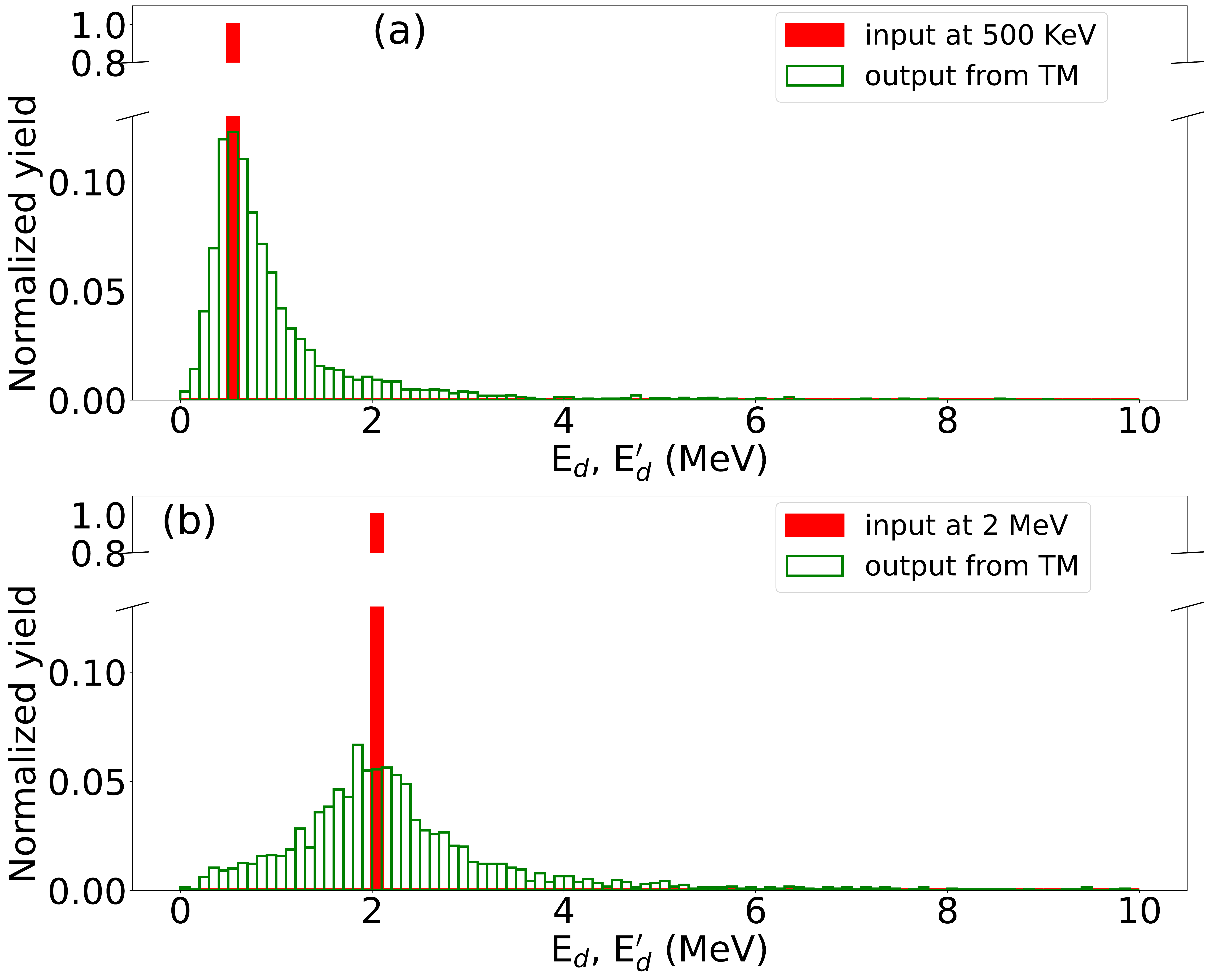}
		\caption{(Color online) Construction of individual columns in the response/transfer matrix (TM). A single bin ($0.2$~MeV width) in input energy $E_d$ is uniformly populated with events, as illustrated by the solid (red) histograms. Processing of the events through a simulation of the detection system yields corresponding event partition across bins in $E_d'$ illustrated by the open (green) histograms.}
		\label{fig:accTM}
\end{figure}

In the experiment considered here, the strongest distortion of the spectrum stems from the acceptance and resolution effects of the Modular Neutron Array and Large multi-Institutional Scintillator Array (MoNA-LISA), and the fact that it is hard to detect neutrons with good efficiency and determine their location with good precision. Detailed simulations of the detector setup allow to quantify the impact of the detection process on a decay energy spectrum and cast it in the form of a response matrix or transfer matrix, cf.~Figs.~\ref{fig:TM} and~\ref{fig:accTM}. The matrix folded with any input decay spectrum and no detector distortions produces the spectrum expected to be measured in the experiment with those distortions imposed.

In constructing the matrix, decays are simulated by randomly drawing the decay energy, $E_d$, from a uniform distribution and randomly selecting the orientation of the decay event in the $^{26}\text{O}$ frame. Each decay is processed through a simulation of the detector response. The decay energy spectrum is then constructed from that response in the same fashion as for the measured data. By selecting a narrow range of input decay energies, the resulting `resolution-folded' spectrum, $E_d'$, for a given $E_d$ is produced (cf.~Fig.~\ref{fig:accTM}).  The $E_d$-values used as examples are shown as thick red lines in Fig.~\ref{fig:accTM} and red arrows in Fig.~\ref{fig:TM}. The full response/transfer matrix is built from the resolution-folded spectra $E_d'$. A difference in normalization for the matrix shown here compared to Fig.~\ref{fig:accTM} should be noted.  There the normalization is for the practical operational and here it is one appropriate for the matrix in continuum limit, representing conditional probability density. As a further note, integration over the measured energy yields the probability of the event at a given input energy getting accepted, $\int dE_d' \, P(E_d'|E_d) = P(E_d)$.  The $E_d'=E_d$ diagonal is marked in the figure to guide the eye.  Tendency to overfill low $E_d'$ values can be observed. The rapid decrease in the probability at high $E_d'$ indicates that an event at high $E_d'$ has a low chance to get recorded.

\vspace{0.25cm}
   \subsection{Accessing resonance properties} 

It is common practice to assess the original, undistorted decay energy spectra with parameter estimation techniques.  For example, neutron-unbound resonances are often \cite{chrisman2021neutron,revel2020f28,leblond2018b21,caesar2013beyond} modeled using energy-dependent Breit-Wigner line shapes~(See Eq.\eqref{Eq31}) \cite{rmatrix}. Parameter estimation methods, such as $\chi^{2}$ minimization, are used to extract the resonance energy,
width 
and angular momentum 
for each resonance state. For the remainder of this paper we refer to such methods as traditional fit methods.

The traditional approaches require decisions on the number of parameters to fit for the original spectrum, such as the choice of the number of resonances present in the explored energy range. The proposed deblurring method aims at restoring the features of the original spectrum without assuming how many states it contains. Using a deep neural network (DNN) classifier method, we attribute probabilities to the hypotheses of different number of states in the original spectrum. When applied to the same data, the two approaches test and complement each other. We complement the results from the two novel approaches by carrying out the standard chi-square minimization and assuming different numbers of resonance peaks in the data.

        \subsubsection{The Richardson-Lucy deblurring procedure}

Our deblurring procedure employs the Richardson-Lucy~(RL) algorithm initially developed to restore blurred images in optics \cite{richardson1972bayesian, lucy1974iterative}. Over time the algorithm found use in astronomy~\cite{thiebaut2016spatially} and medicine for medical images analysis \cite{al2015deblurring}, to list a few. In high-energy physics, analogous developments progressed~\cite{dagostini_multidimensional_1995} without realization of the prior work elsewhere. Recently, Danielewicz and Kurata-Nishimura \cite{danielewicz2022deblurring} have demonstrated that a nonlinear extension of the algorithm could be used to determine three-dimensional (3D) momentum distributions of products in intermediate-energy heavy-ion collisions. The RL algorithm derivation relies on the Bayes' theorem and it follows an iterative procedure to find a self consistent solution. The algorithm only uses the distorted spectrum and discretized response function of the apparatus, or Transfer Matrix (TM), as inputs.  The spectrum entries and matrix elements are positive definite and carry probabilistic interpretation. The restoration of the original spectrum is an inverse problem, but it progresses in the deblurring without directly inverting the TM, an uncommon approach for inverse problems~\cite{grech2008review}. In maintaining the restored spectrum positive throughout the iteration procedure and by avoiding a direct TM inversion, serious singularity problems plaguing inverse problems are avoided.

In Ref.~\cite{danielewicz2022deblurring}, the RL was implemented without consideration of noise. In the present work, we expand the utility of the algorithm by considering measurement statistics and improve on the assessment of what is actually learned from the data.
However, in other fields, it has been demonstrated that the RL algorithm suffers from short-wavelength instability due to noise amplification after a limited number of iterations \cite{fister2007deconvolving, dey2006richardson,vargas_unfolding_2013}. To overcome this challenge, we introduce a regularization in the algorithm that tames the short wavelength component in the deblurring solution. There are several options for such regularization, the Gaussian function smoothing being one such example.  The smoothing requires considerations of a function width and boundary conditions \cite{fister2007deconvolving}.  Another regularization option is the use of denoising algorithms that invoke nonlinear combinations of derivatives of restored {spectra}~\cite{dey2006richardson,rudin1992nonlinear}, commonly termed Total Variation (TV).

In this work, we use a simple version of TV regularization employed in Ref.~\cite{danielewicz2022deblurring}, but we make its strength increase with energy, as the impact of noise on a restored spectrum increases at higher energy. With this approach we are able to arrive at stable deblurring solutions after just few RL iterations.
 
\subsubsection{The Deep Neural Network classification algorithm}
            
In addition to the RL based deblurring algorythm, we implemented a Deep Neural Network (DNN) classification algorithm in our analysis procedure to identify the number of resonance states in the decaying nucleus (i.e., $^{26}\mathrm{O}$) from the measured decay energy distribution. 
The DNN methods have been popular in face~\cite{554195} and speech~\cite{8632885} recognition. In the field of particle and nuclear physics, the methods have been applied to particle identification and event selection \cite{guest2018deep, matchev2021thickbrick, carleo2017solving, whiteson2009machine,fujimoto2020mapping, bedaque2021ai}.
In the present work, the DNN uses a training dataset generated from a Breit-Wigner (BW) resonance distribution, folded with the experimental response matrix and sampled according to a Poisson distribution. This process yields a dataset which resembles experimental data. 
The dataset is labeled and grouped into classes based on the number of resonance peaks introduced in the BW distribution. 

\section{Deblurring Algorithm}

\subsection{Setting}
In a nuclear decay experiment, the decaying nucleus, characterized by a total four momentum 
$\mathbf{p}=(E, p_x, p_y, p_z)$, can be thought of as an emitter of particles that fly off towards the detector. The detector records the particles with some efficiency and allows to determine their four-momenta with some accuracy. From the combination of those four-momenta, the invariant mass of the decaying nucleus is determined, $M=\sqrt{\mathbf{p}^2}$, and, over many events, the particle decay energy spectrum is established~\cite{redpath2019measuring, redpath2020new}. Structures in that spectrum can tell us about the resonance states of the decaying nucleus. Limitation in the detector resolution makes the measured spectrum $f$ blurred compared to the true decay spectrum $\mathcal{F}$ of the nucleus. 

The blurring relation between $f$ and $\mathcal{F}$ can be written as
\begin{eqnarray}
f(E_d^{\prime})=\int dE_d \, P(E_d^{\prime}|E_d) \, \mathcal{F}(E_d) \,  .
\label{Eq1}
\end{eqnarray}
Here, $E_d'$ is the measured energy, $E_d$ is the true energy and $P(E_d^{\prime}|E_d)$ is the conditional probability that products for a nucleus decaying at $E_d$ are registered, the event is accepted and determined to represent the decay energy $E_d'$. 
In the context of an experiment, $P(E_d^{\prime}|E_d)$ represents the response function of the apparatus, but in the context of blurring analyses it may be called a blurring or transfer function. As an extreme example, $P(E_d^{\prime}|E_d) = \delta(E_d^\prime - E_d)$ represents an ideal detector.

Eq.~\eqref{Eq1} invokes the spectra $f$ and $\mathcal{F}$ in the limit of infinite measurement statistics.  In practice, the spectra get discretized, most often simply binned. Moreover, in an experiment, $f$ only gets determined with some accuracy, and even $P$ gets established with some resolution.  Under discretization, the blurring relation \eqref{Eq1} acquires the matrix form
\begin{eqnarray}
f_i=\sum_j P_{ij} \, \mathcal{F}_j \, ,
\label{Eq11}
\end{eqnarray}
where $1 \le i \le N$, $1 \le j \le M$ and $P_{ij}$ represents the conditional probability density integrated over a discretization form factor (typically $\Delta E$ bin) in $E_d$ and averaged over one in $E_d'$.  As such, the matrix elements $P_{ij}$ are positive and $P_i = \sum_j P_{ji}$ represents probability than an event at decay energy near $E_d^i$ is analysed.  

In our analysis of decay-energy spectra, we most often employ $\Delta E_d = 0.2$~MeV binning. To construct the transfer matrix (TM), $P(E_d^{\prime}|E_d)$, for the three-particle decay $\nuc{26}{O} \rightarrow \nuc{24}{O} + 2n$ experiment \cite{redpath2019measuring}, we randomly draw the decay energy, $E_d$, from a uniform distribution and draw the orientation of the decay event in the frame of $^{26}\text{O}$. Each decay is then processed through the simulated response  \cite{redpath2019measuring, redpath2020new} of the detector setup schematically illustrated in Fig.~\ref{fig:sweeperconcept}. The outcomes are sorted by bins in $E_d$ and $E_d'$ and their counts per $E_d$ bin entry become TM elements.  The constructed matrix is illustrated in Fig.~\ref{fig:TM}.  The TM construction is additionally illustrated in Fig.~\ref{fig:accTM} for individual $E_d$ bins. An $E_d$ bin is uniformly populated with events, as indicated by the solid (red) histograms shown in Fig.~\ref{fig:accTM}.  Those events are processed through the simulation of the detector response and sorted according to~$E_d'$ bins, as indicated by the open (green) histograms.  After renormalization, the open (green) distributions in Fig.~\ref{fig:accTM} become columns in the TM normalized as probability density $P(E_d^\prime|E_d)$, or as contributions to the probability $P_{ij}$ in practical calculations with discretized spectra.

\subsection{Deblurring}

The goal of deblurring is to estimate $\mathcal{F}$ when only~$f$ and $P$ are known.  The Richardson-Lucy (RL) algorithm \cite{richardson1972bayesian, lucy1974iterative, dagostini_multidimensional_1995, danielewicz2022deblurring} relies on the conditional probability $Q(E_d  |E_d')$ complimentary to $P(E_d'| E_d)$.  Bayesian theorem linking the two probability densities yields a set of equations \cite{danielewicz2022deblurring} that can be solved for $\mathcal{F}(E_d)$ by iteration:
\begin{eqnarray}
f^{(n)}_j &= &\sum_i P_{ji} \, \mathcal{F}^{(n)}_{i} \, ,
\label{Eq19}
 \\
\mathcal{F}_i^{(n+1)} & = & \mathcal{F}^{(n)}_i \, \sum_j \frac{f_j}{f^{(n)}_j} \, \frac{P_{ji}}{P_i} \, .
\label{Eq20}
\end{eqnarray}
Here, $n$ is the iteration index.

We have chosen to start RL iterations with a rough guess for $\mathcal{F}^{(0)}$, such as scaled up $f$. The iterations is stopped once $\mathcal{F}^{(n)}$ ceases to change with $n$.  For distributions that quickly change with their arguments, such as $E_d$ here, the long-term convergence may be slow and for large $n$ numerical seesaw instabilities in the arguments may set in.  That instability can be tamed with a renormalization factor \cite{dey2006richardson, danielewicz2022deblurring} $I^{(n)}$ applied to the r.h.s.\ of \eqref{Eq20}:
\begin{eqnarray}
I^{(n)}=\frac{1}{1 - \lambda \, {\mathbf{D}} \cdot {\boldsymbol \nabla} \Big(\frac{{\boldsymbol \nabla} F^{(n)} }{|{\boldsymbol \nabla} F^{(n)}|} \Big)} \, .
\end{eqnarray}
Here, $\mathbf{D}$ is a vector with components that are intervals over which $\mathcal{F}$ is discretized in its arguments (bin sizes), the divergence is approximated in low order based on that discretization and $\lambda$ is a small positive number.  In a one-dimensional case, such as here, the factor becomes simply
\begin{eqnarray}
I_i^{(n)}=
\begin{cases}
\frac{1}{1-\lambda}  \, , & \text{if} \, \, \mathcal{F}^{(n)}_i < \mathcal{F}^{(n)}_{i-1,i+1} \, , \\
\frac{1}{1+\lambda}  \, , & \text{if} \, \,  \mathcal{F}^{(n)}_i > \mathcal{F}^{(n)}_{i-1,i+1} \, , \\
1 \, , & \text{otherwise} \, .
\end{cases}
\label{filter}
\end{eqnarray}
This factor suppresses any patterns of maximae and minimae emerging on the discretization scale. However, when wider-scale maximae or minimae arise, the factor will be impacting them too. As uncertainties in the restored $\mathcal{F}$ will be of interest here, the use of the above regulation factor will introduce a relative error of the order of $\lambda$ around the extrema of the restored~$\mathcal{F}$.

\subsection{Fluctuations and other practicalities}

The blurring relation \eqref{Eq1} invokes spectra in the limit of infinite statistics.  However, the spectra are measured at finite statistics and its characteristics are expected to fluctuate compared to those at high statistics.  

Let $f$ represent the average event numbers registered in different bins of decay energy for measurement series carried out over a specific measurement time. If we carry out repeated measurement series over that time, event counts for individual bins will fluctuate in a Poisson-like manner. If we carry out just one measurement series, then the event count in the bin $i$, $f_i$, is our best estimate for the mean count and the best estimate for the mean squared deviation from that mean over repeated series \cite{bohm_introduction_2010}.

When assessing uncertainties in the restored spectra, we build up an ensemble of alternative measurement results over the same time, consistent with the best estimates of the mean values for decay energy bins and dispersion, by sampling the Poisson probability distribution for content $f_i^*$,
\begin{eqnarray}
\mathcal{P}(f^*_i|f_i)=\frac{e^{-f_i}f_i^{f^*_i}}{f^*_i!} \, .
\label{Poisson}
\end{eqnarray}
We then carry out the RL restoration, Eqs.~\eqref{Eq19}-\eqref{Eq20}, with $f_i$ replaced by $f_i^*$, arriving at $\mathcal{F}^*$ and we study the distribution of the latter within the ensemble. The algorithm requires  $\mathcal{F}^{(0)}\geq 0$ to start. However, we have not seen any significant sensitivity of the results to the fine details of $\mathcal{F}^{(0)}$. In practice, the important factor is the number of iterations, a few hundreds is sufficient in our case, and the smoothing factor~(see Eq.~\eqref{eq:EdepSmoothing}).

Within the higher end of the decay energy window in which we operate, usually up to $10 \, \text{MeV}$, the counts tend to be low, fluctuating with energy and these fluctuations tend to be amplified in the restoration. Correspondingly, we make the parameter $\lambda$ in the factor $I$, Eq.~\eqref{filter}, increase with energy:
\begin{eqnarray}
\lambda=\lambda_0 \, \Big(1+\Big(\frac{E}{E_0}\Big)^2 \Big) \, ,
\label{eq:EdepSmoothing}
\end{eqnarray}
and we typically use $\lambda_0=0.035$ and $E_0 = 6 \, \text{MeV}$.  The form and parameter values have been adjusted through experimentation. Notably, $\lambda$ increases the bin-to-bin correlation, which is illustrated in Fig.~\ref{fig:Corr}.
The values for $\lambda _0$ and $E_0$ are chosen to reduce the noise oscillations in the restoration; this could also be done by increasing bin sizes. 
We choose $\lambda$ to depend on E to suppress  the oscillations in the restoration in the high E range, which is due to the finite statistics.

\section{Tests of deblurring algorithm \label{sect:3}}
In this section we carry out tests of our deblurring procedures when applied to simulated data.  We first consider data with negligible errors and then data with statistical errors comparable to those for the investigated decay energy measurements \cite{redpath2019measuring,redpath2020new}.

Following physical expectations regarding the forms of input decay-energy spectrum, the spectrum $\mathcal{F}(E_d)$ in the tests is modeled as a superposition of Breit-Wigner distributions:
\begin{eqnarray}
\mathcal{F}(E_d)\approx \sum_i A_i \, \frac{0.5 \, \Gamma _i}{(E_d-E_i)^2+(0.5 \, \Gamma_i)^2} \, .
\label{Eq31}
\end{eqnarray}
\begin{figure}[H]
    \centering
    \includegraphics[scale=0.4]{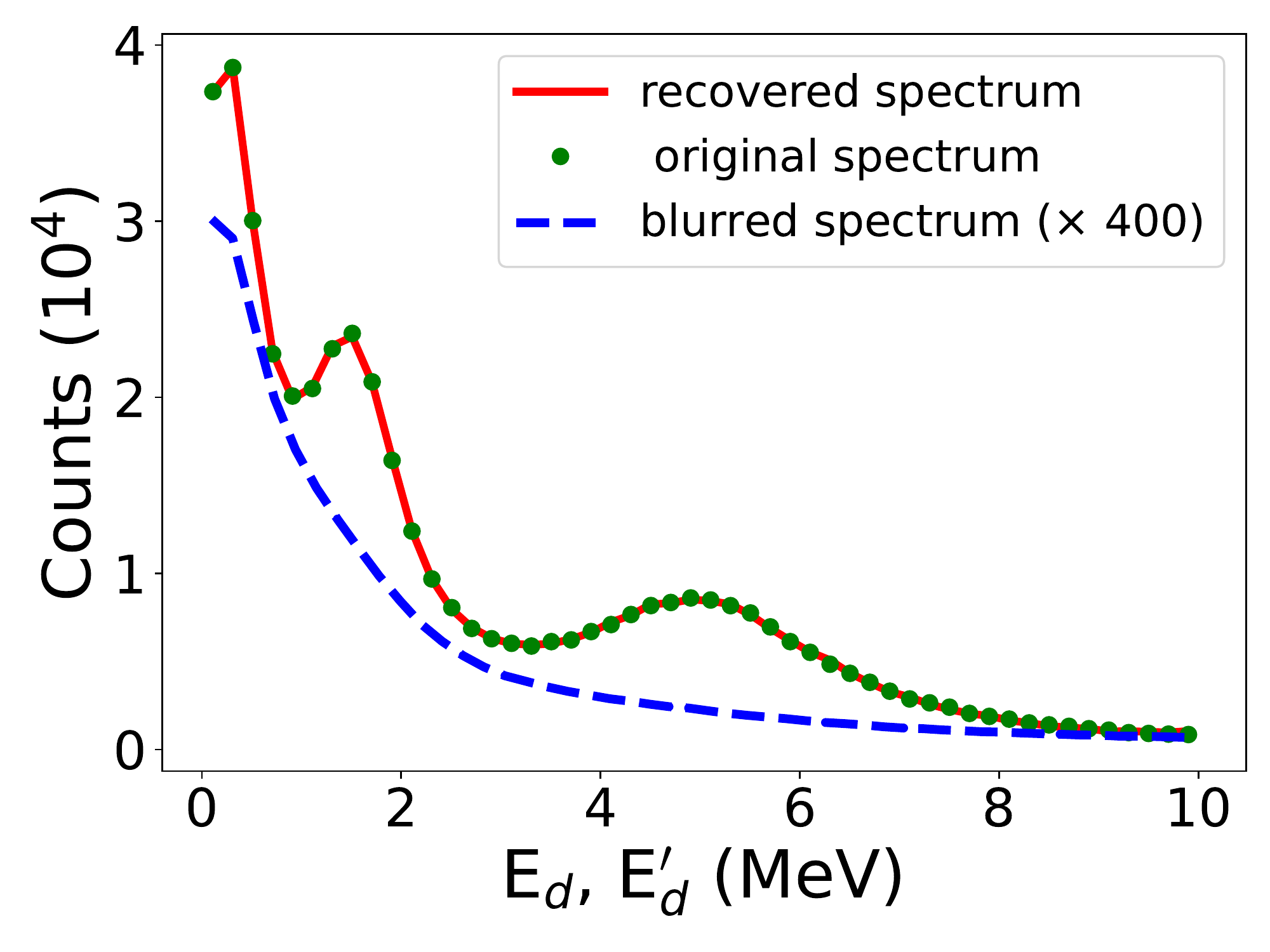}
    \caption{(Color online) Restoration of decay-energy spectrum in the absence of noise. The dots (green) represent the original event distribution modelled with Eq.~(\ref{Eq31}). Three wide peaks were assumed for the spectrum. The dashed (blue) line represents the blurred distribution, at adjusted normalization, and it has been obtained by folding the original distribution with the TM, cf.~Eq.~\eqref{Eq11}. The solid (red) line represents the distribution obtained by subjecting the blurred spectrum to deblurring with the RL algorithm, Eqs.~\eqref{Eq19} and \eqref{Eq20}. The restored and original distributions lie practically on top of each another. A binning in energy of $0.2 \, \text{MeV}$ was employed in generating these spectra.  
       }
    \label{fig:my_label}
\end{figure}

We are generally interested in the decay energy region extending up to $10 \, \text{MeV}$, though we have also considered energies up to $14 \, \text{MeV}$. Within such regions we have experimented with distributions containing (1--5) Breit-Wigner peaks at different energies $E_i$ and of different widths $\Gamma_i$ and amplitudes $A_i$.  In the case we will use here for illustration, we take three peaks at~$0.3$, $2$, and $4.5 \, \text{MeV}$, with respective widths of $0.3$, $0.85$, and $1.3 \,  \text{MeV}$, see Fig.~\ref{fig:my_label}. For simplicity, we take $A_i \equiv 1$. 
\begin{figure*}
    \centering
    \includegraphics[scale=.18]{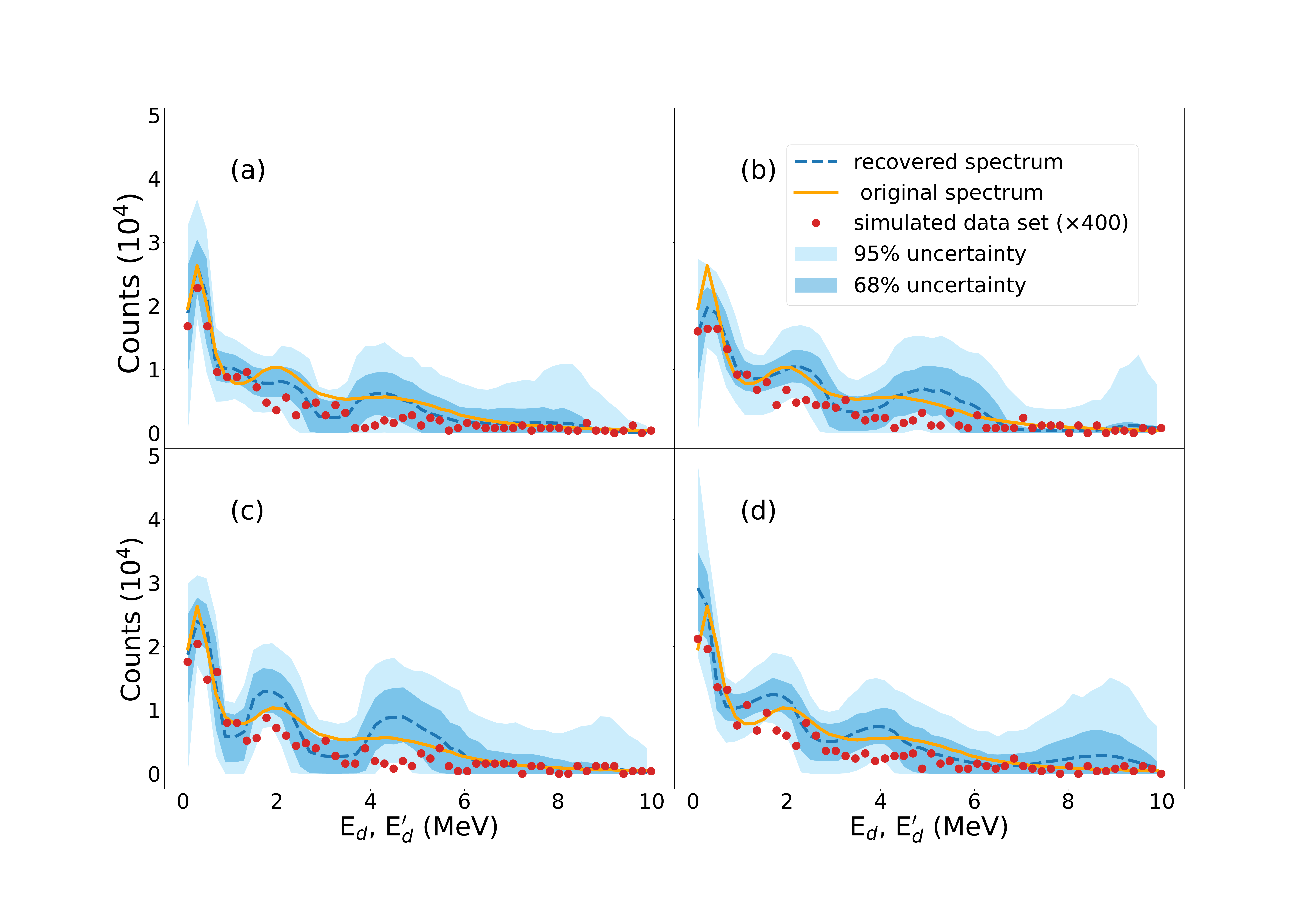}
    \caption{(Color online) Restoration for four different examples of simulated data sets when Poisson noise is active.  The sampled original spectrum is the same for each set and event statistics behind each set is similar to that believed to be behind the real data analysed in this work. The points represent the individually sampled sets with counts scaled up by a factor of 400. The  dark blue and light blue bands illustrate the $\sigma$ and $2\sigma$ uncertainties resulting from spectra restoration with error sampling. In each panel, the dashed blue line represents the mean in the restoration ensemble for the set. The original spectrum (solid orange line) has three resonance peaks located  at~$0.3$, $2$, and 4.5 MeV with respective widths of $0.3$, $0.85$, and 1.3 MeV.  Generally, we succeed in restoring the structures in the original spectrum using the RL algorithm. Binning of 0.2 MeV was used for the processed spectra. 
    }
    \label{fig:my_label1}
\end{figure*}
At first, we take the modeled input distribution~$\mathcal{F}$ and multiply it by the TM to get $f$.  Up to some joint normalizing factor for both, these distributions stand for those in the limit of a very large statistics.  The simulated input and measured distributions are illustrated in Fig.~\ref{fig:my_label}. To the simulated measured distribution we apply the RL algorithm, Eqs.~\eqref{Eq19} and \eqref{Eq20}.  The restored distribution from the iteration is also shown in Fig.~\ref{fig:my_label} and it lies practically on top of the original. This has been our typical finding for the limit of large statistics, no matter what input.  In the restorations for large statistics, we usually can drop the smoothing factor \eqref{filter}.

Next, we turn to simulations of ensembles of events, such as for real data.  Specifically, we sample the shape of~$\mathcal{F}$ within the energy range $E_d < 10 \, \text{MeV}$, to get $E_d$ for a single event.  Then we sample the probability density $P(E_d'|E_d)$ from TM to decide whether this event is accepted for analysis and what the measured $E_d'$ is. We repeat the process until the number of analysed events is similar to that in the experiment.  The needed number of input events provides a normalization for $\mathcal{F}$. In Fig.~\ref{fig:my_label1}, we show results from such four separate data simulations. Both the simulated measured $f(E_d')$ and underlying $\mathcal{F}(E_d)$ are shown there.

A measurement carried out over a specific beam time, with finite statistics, can be viewed as a member of an ensemble of measurements ran over the same time. We next attempt to simulate such an ensemble using only information in an individual generated data set, following the Poisson distribution sampling discussed earlier, Eq.~\eqref{Poisson}, to get $f^*(E_d')$. To the individual $f^*$, we apply the RL deblurring algorithm to get an estimate of $\mathcal{F}$.  With this, we arrive at an ensemble of restored $\mathcal{F}$ that reflects uncertainties inherent in $f$, within the methodology we adopt. In Fig.~\ref{fig:my_label1}, we further show the characteristics of the ensemble of restored $\mathcal{F}$, for each simulated data set, specifically the average values for the bins and 68\% and 95\% uncertainty ranges. It can be observed that the distributions of the restored values are generally consistent with the input $\mathcal{F}$.  

\begin{figure}[H]
    \centering
    \includegraphics[width=\linewidth]{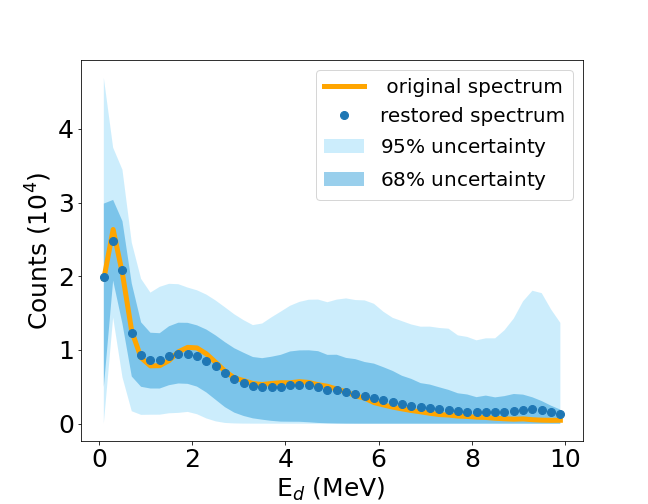}
    \caption{(Color online) Outcome of averaging over restored distributions from $24$ such simulations as in Fig.~\ref{fig:my_label1}. The overall mean (dots) and the original spectrum (solid) are largely on top of each other. 
        }
    \label{fig:my_label2}
\end{figure}

We complement the above resampling results by showing in Fig.~\ref{fig:my_label2} a distribution of restored $\mathcal{F}$ resulting from averaging over the distributions of restored $\mathcal{F}$ from a number of individual data simulations such as in Fig.~\ref{fig:my_label1}. It can be seen that the average over a large number of ensembles begins to approach the input $\mathcal{F}$ suggesting a faithful nature of the restored $\mathcal{F}$ even for finite statistics at the level of smoothness expected for decay spectra and accuracy that may be aimed at currently.
\begin{figure}[H]
    \centering
    \includegraphics[scale=0.4]{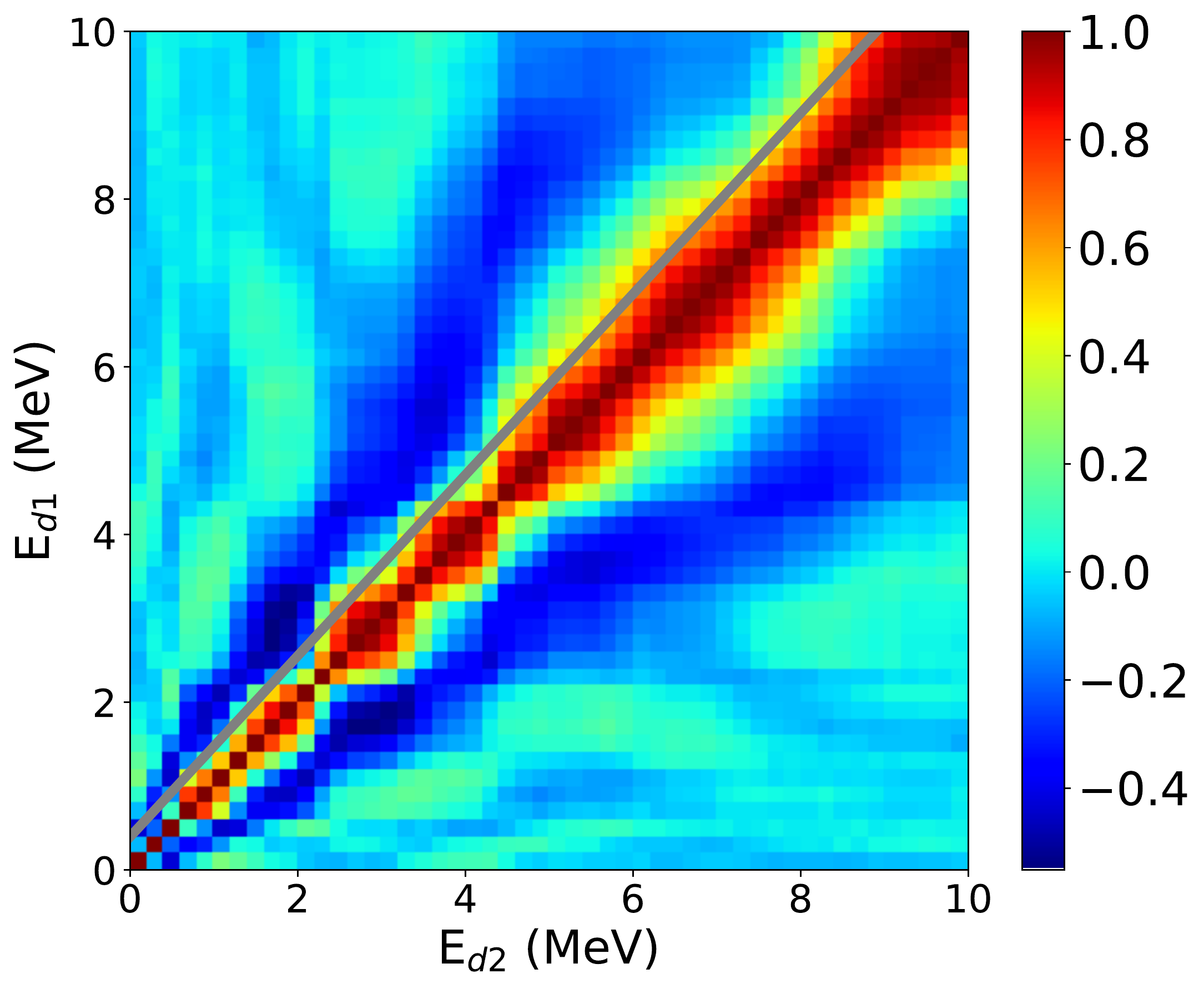}
    \caption{(Color online) Contour plot of bin to bin Pearson correlation matrix for the restored spectrum when carrying out resampling for the case of the simulated spectrum in Fig.~\ref{fig:my_label1}(d). The solid line guides the eye to indicate average behavior of the width for the main peak in the correlation -- the finer details with energy can depend on the assumed original spectrum and even particular simulation.  On average, the width grows with energy.}
    
    \label{fig:Corr}
\end{figure}

The TM with binning for the measured decay energy as well as the RL algorithm with smoothing will generate correlations in results for different bins in the restored energy spectrum.  Such correlations can limit the resolution that one can aim at for the restored spectrum.  In resampling, we can test the emergence of the inter-bin correlations.  This is demonstrated in Fig.~\ref{fig:Corr} which shows bin to bin Pearson correlation matrix built from the restored spectrum shown in Fig.~\ref{fig:my_label1}(d). A solid line in the figure guides the eye to show the average behavior for the width of the main peak in the correlation.  Beyond variation tied to specific assumptions on the underlying spectrum, the width generally increases with the decay energy, starting at about $0.4 \, \text{MeV}$ at low $E_d$ and rising to $1.1 \, \text{MeV}$ at $E_d \sim 10 \, \text{MeV}$.

\section{Deblurring $^{26}$O decay energy spectrum }

In the experiment \cite{redpath2019measuring,redpath2020new}, two-neutron unbound \nuc{26}{O} was produced via one-proton knockout from a \nuc{27}{F} beam.  The \nuc{26}{O} nucleus decayed to $\nuc{24}{O} + n +n$, and position and time-of-flight measurements of the daughter products were carried out in order to assess their momenta. The momenta for \nuc{24}{O} and two neutrons, measured in coincidence, were used to reconstruct the decay energy spectrum for \nuc{26}{O} using the invariant mass technique.

\begin{figure*}
	\centering
	\includegraphics[width=.71\linewidth]{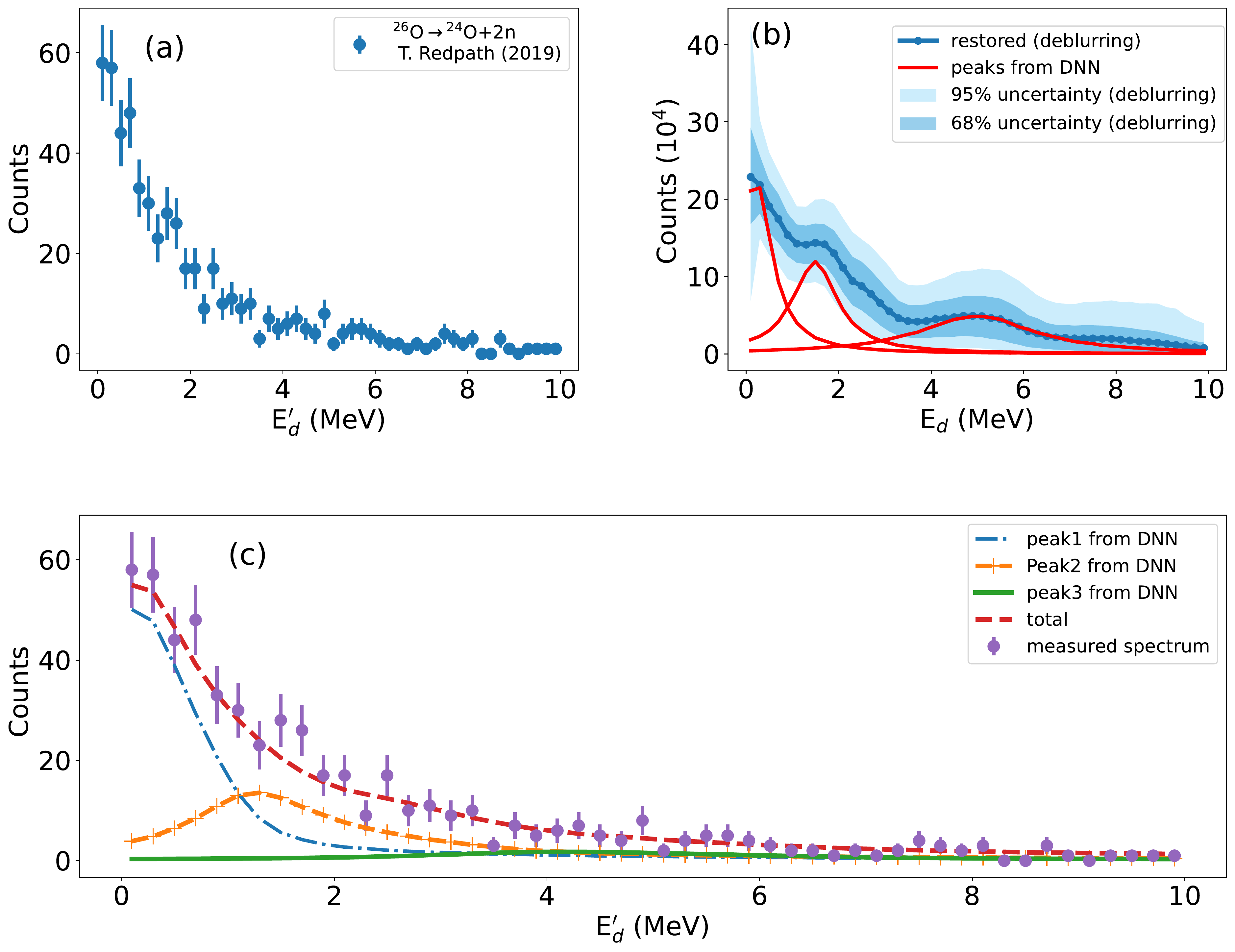}
	\caption{ (Color online)  Analysis of the measured three-body decay energy spectrum for ${}^{26}\mathrm{O} \rightarrow {}^{24} \mathrm{O} + 2n$.  Panel (a) shows the spectrum measured using invariant mass spectroscopy \cite{redpath2019measuring,redpath2020new}. Panel (b) shows the deblurred spectrum, as well as the peaks identified for the spectrum with the Deep Neural Network (DNN).  Resonances behind the peaks in the spectrum near 0 and 1.3 MeV were also identified for $^{26}\mathrm{O}$ in Ref.~\cite{kondo2016nucleus}  (0$^{+}$ and 2$^{+}$ states, respectively). Indications of a third peak at about 4 MeV were reported by Caesar {\em et al.}~\cite{caesar2013beyond}. Panel (c) displays contributions from the three peaks identified by DNN, and shown in (b), to the measured spectrum, i.e., after blurring caused by the apparatus. Combination of those contributions (dashed line) matches closely the data (points). The width of the energy bin in processing the spectra is 0.2~MeV. 
	}
  \label{fig}
\end{figure*}


Previous invariant mass measurements have observed the ground state of $^{26}$O decaying directly into $^{24}\text{O}$ and neutrons very near threshold \cite{kohley2013study,kondo2016nucleus,caesar2013beyond,redpath2019measuring}. A recent experiment measured the half-life of this state to be $5\, \text{ps}$~\cite{redpath2020new}. An excited state, $^{26}$O(2$^+$), was also measured with a decay energy of $1.28 \, \text{MeV}$ above threshold~\cite{kondo2016nucleus}.
Indications of a high-lying excited state, at around $4 \, \text{MeV}$, were reported in Ref.~\cite{caesar2013beyond}, but Ref.~\cite{kondo2016nucleus} found no evidence of that state. 
\begin{figure}
    \centering
    \includegraphics[scale=0.5]{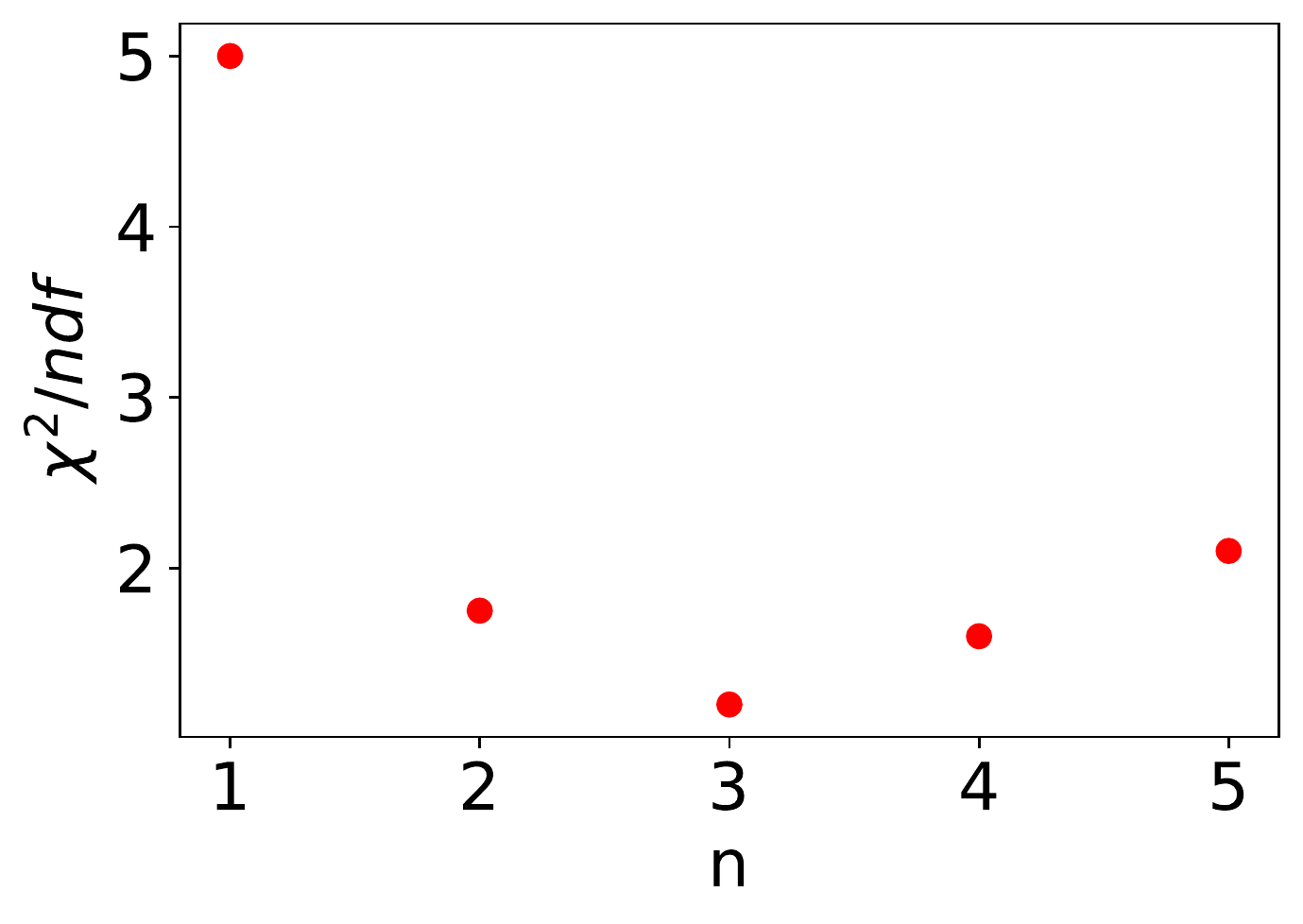}
    \caption{ (Color online) Chi-square per degree of freedom versus the number of peaks included in the fit to the experimental decay energy spectrum (shown in Fig.~\ref{fig}(a)). The horizontal axis represents the number of peaks, $n$, and each peak is described by three parameters (see the text for details). Increasing the number of peaks is equivalent to increasing the number of parameters for fitting. It may be seen that the three-peak case yields the minimal chi-square per degree of freedom.}
    \label{chisqrt}
\end{figure}
Panel (a) of Fig.~\ref{fig} shows the energy spectrum of the three-body decay of $^{26}$O as recorded in the experiment performed at NSCL \cite{redpath2019measuring}. Only the first peak, from those mentioned above, is easily seen.
The deblurring technique discussed in the previous section helps to extract more information from the measured decay energy spectrum. Panel (b) of Fig.~\ref{fig} displays the spectrum restored from the measured spectrum of the $^{26}$O system, using the deblurring method, Eq.~(\ref{Eq20}) with an energy-dependent smoothing parameter of Eq.~\eqref{eq:EdepSmoothing}. The bumps evident in the restored spectrum  near $0 \, \text{MeV}$ and $1.3 \,  \text{MeV}$, respectively, can be recognized as the $J=0^+$ and $2^+$ states of $^{26}$O nucleus identified in Ref.~\cite{kondo2016nucleus}. We associate the broad peak between 4 and $6$~MeV with the third $^{26}\mathrm{O}$ state observed in \cite{caesar2013beyond}. 
The panel (b) in Fig.~\ref{fig} includes peaks that DNN attributed to the original spectrum and corresponding contributions of those peaks to the observed spectrum. We discuss the DNN analysis of decay energy spectrum next.

In comparing our method with traditional methods, we have performed chi-square minimization  by fitting the measured decay energy spectrum with the resolution-folded BW distribution (see Section. IV). We started with one peak BW function, and gradually increased the number of peaks to five. Each peak is described by three parameters, i.e., amplitude, peak position, and peak width, which implies that the number of fit parameters is three times the number of peaks. In Fig.~\ref{chisqrt}, we present the values of chi-square per degree of freedom, $\chi^2/ndf$, versus the number of peaks,~$n$. A decrease in $\chi^2/ndf$ may be observed from $n=1$ to $n=3$ and then an increase from $n=3$ to 5, which implies that three peaks are sufficient to describe the data.

It is important to emphasize that the deblurring method does not require any assumption about the number of peaks in the spectrum in order to carry out the restoration, whereas, in the chi-square approach as well as DNN (to be discussed in the next section), one needs to invoke some peaks explicitly (or parameters) in the model. From its side, the deblurring method can suggest the type and number and type of parameters needed in the chi-square fitting or DNN. 

\section{DNN architecture to discover resonance states}

\begin{figure}[H]
\centering
	\includegraphics[scale=0.35]{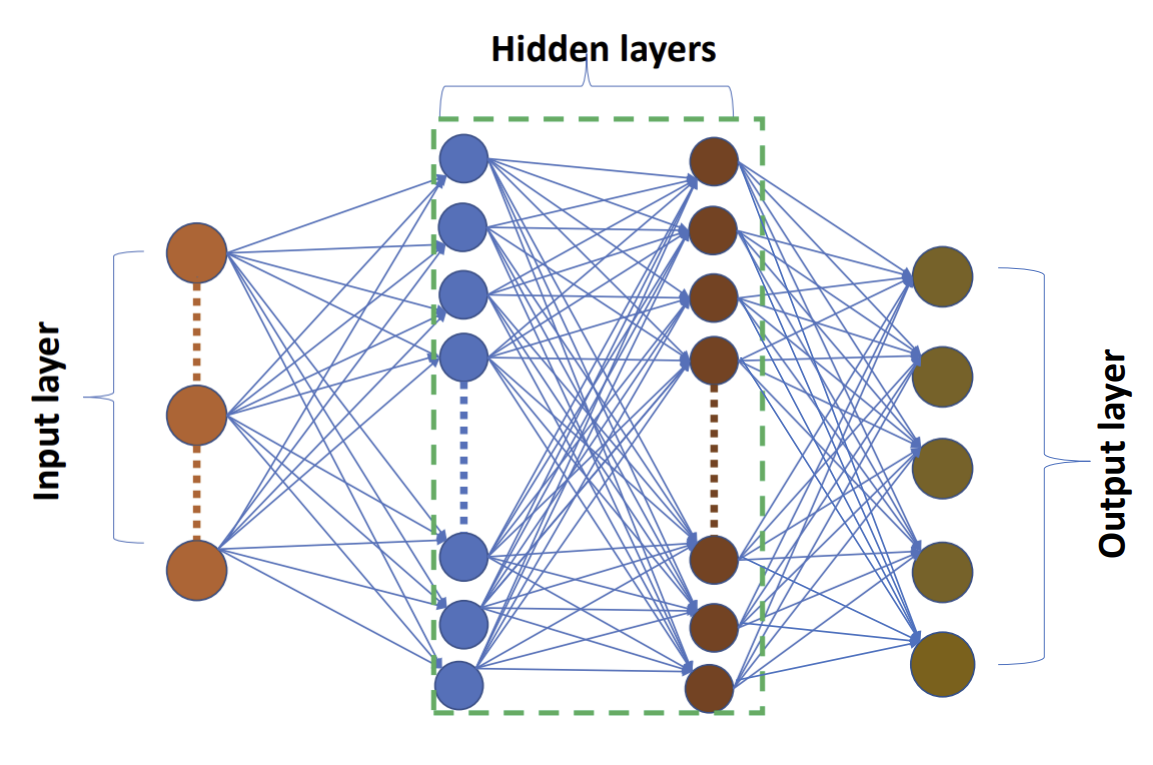}
	\caption{ (Color online) The figure shows a schematic illustration of deep neuron network architecture designed for the classification model. In the present work the input (at the input layer) is the decay energy spectrum, and at the output layer, is a labeled (class) value which tells the number of states in the spectrum. The parameters used to train and test the model are displayed in Table~\ref{tab:table1}.}
	\label{DNN}
\end{figure}

 Alongside the deblurring method, we built a machine learning (ML) tool to classify the number of peaks in the observed decay energy spectrum. A fully connected DNN, schematically illustrated in Fig.~\ref{DNN}, is defined with the equations:
\begin{eqnarray}
A^{l+1}_i&=&b^{l}_{i}+\sum_{j+1}^{(l)}W_{ij}a_j^{(l)},\\
a_j^{(l)}&=& Z(A^{l}_i).
\end{eqnarray}
where $a^{(l)}$ and $A^{l+1}$ are the input and output layers and $W_{ij}^{l}$ and $b^{l}$ are the weights and bias of the $l^{st}$ layer. The non-linear activation function is $Z(x)=\text{ReLu}=\rm{max}(0,x)$.  The Relu \cite{daubechies2019nonlinear} is commonly used as the activation function in neural network models. The function $f(x)_i=\rm{Softmax}=\frac{e^{x_i}}{\sum_i^Ne^{x_i}}$ is used in the output layer to normalize or scale the output so that it may be interpreted as a probability~\cite{sharma2017activation}. We implement the network using the categorical cross-entropy loss function,
$L=-\sum_{i}^{N} y_i\log(\tilde{y}_i)$, that is suitable for a multi-class classification problem \cite{rusiecki2019trimmed}.  Here, $y_i$ is the $i^{th}$ actual value and $\tilde{y}_i$ is the $i^{th}$ predicted value (output of the DNN).
Then, the Adaptive Moment Estimate (Adam) algorithm \cite{kingma2014adam}, a popular optimizer in DNN models, is used to solve for the optimal weights $W_{ij}$. The architecture and training specifications of the DNN model are displayed in Table~\ref{tab:table1} and the network design is shown in Fig.~\ref{DNN}.

The DNN classifier is trained using simulated datasets to learn plausible patterns in the decay energy spectra. The dataset is simulated by by folding Breit-Wigner line shapes, Eq.~\eqref{Eq31},  with the TM in order to resemble the experimental spectra. We then distort the folded distribution according to Poisson noise to produce a noisy distribution similar to experimental measurements. The parameters $E_i$ and $\Gamma_i$ in Eq.~\eqref{Eq31}, with $i=1,\ldots, 5$, are randomly drawn from a uniform distribution. In this work, we consider the parameters to stem from the range of values displayed in Table~\ref{tab:table1}.

We divided the training data set into five classes of spectra according to the number of resonances contributing to the decay energy spectrum.
The first class, C$_1$, assumed two resonance states with energies $E_1$ and $E_2$. The second class, C$_2$, assumed three resonances at $E_1$, $E_2$, and $E_3$. The third class, $C_3$, assumed three resonances at $E_1$, $E_2$, and $E_5$. The fourth class, C$_4$, assumed four resonances at $E_1$, $E_2$, $E_3$, and $E_4$. The class C$_5$ contained any other spectrum that does not belong in the first four classes. For convenience, we assign to class C$_5$ four kinds of spectra: spectra with one peak at E$_0$, spectra with two peaks at E$_0$ and E$_2$, spectra with three peaks at E$_0$, E$_3$, and E$_4$ and spectra with four peaks at E$_0$, E$_1$, E$_2$, E$_3$, and E$_4$. It is important to note that, in choosing values for $E_{1,2,3,4}$, we made sure to include all the $^{26}$O states that were previously reported (see Refs.~\cite{kondo2016nucleus,caesar2013beyond}). The mean values of $E_{1,2,3,4}$ have been equal to about 0.15, 1.50, 2.40, and 5.00~MeV, respectively.  

\begin{figure}[H]
	\includegraphics[scale=0.35]{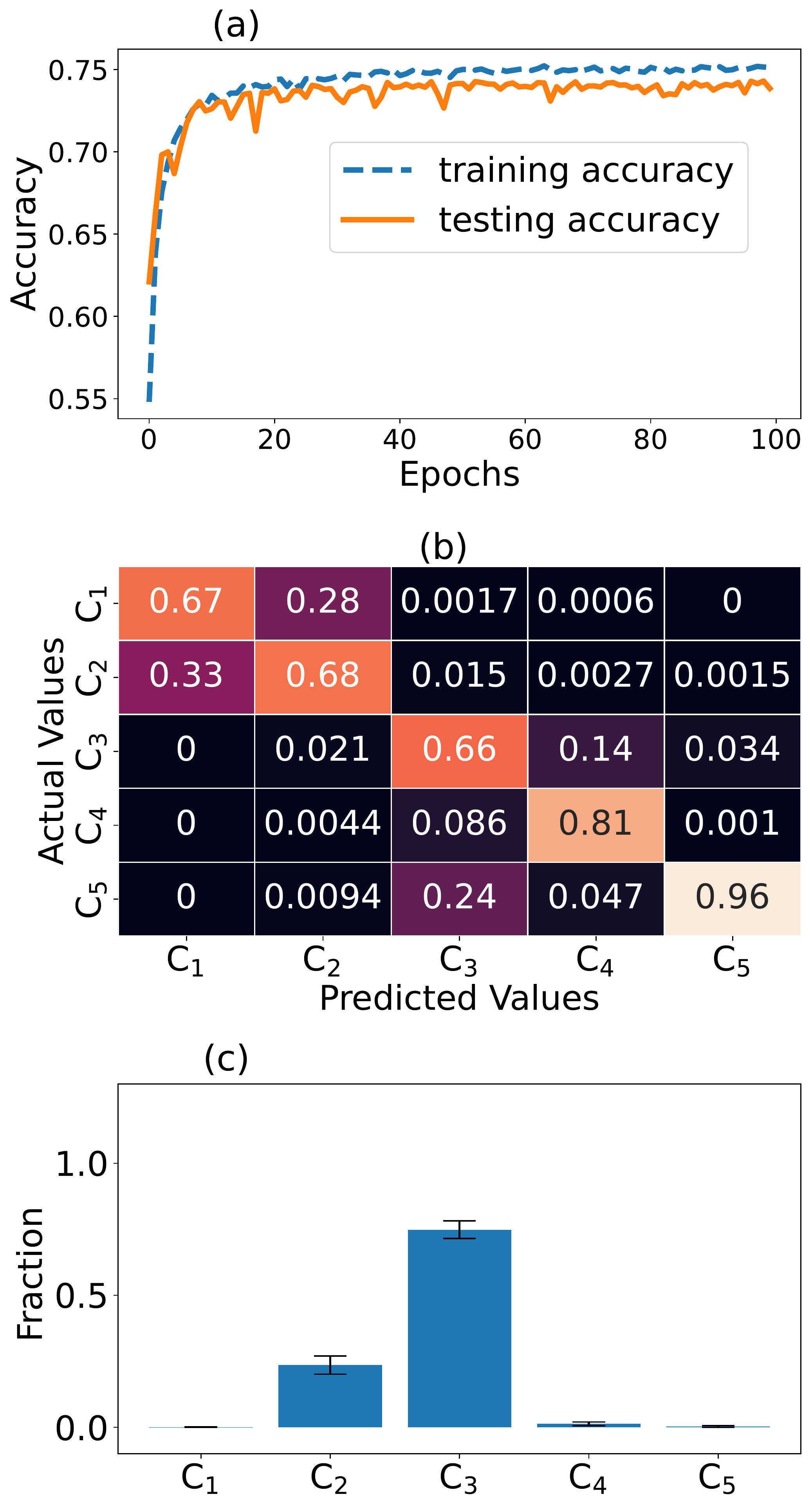}
	\caption{(Color online) DNN model to identify resonance states from measured decay energy spectrum. Panel (a) indicates the training and testing accuracy of the model. The curves converge at $\approx$ 0.75 on both data sets, which means the model predicts 75\% of the data set correctly. Panel (b) represents the confusion matrix which tells how well the DNN classifier was able to classify spectra: C$_1$, C$_2$, C$_3$, C$_4$, and C$_5$ are five classes we used to train the model, and the detail about each class is discussed in the text and Table.\ref{tab:table1}. Panel (c)  displays the ratio of distributions predicted to belong in a given category (C$_i$, $i=$1, 2, 3, 4, 5) over the total number of distributions used in the DNN model prediction. The ratio helps to estimate the class where the measured spectrum fits. We found that C$_3$ has the highest fraction, which suggests that there is a high chance for the experimental decay energy spectrum of $^{26}\mathrm{O}$ system decaying $^{24}\mathrm{O} +$ 2n from invariant mass spectroscopy measurements belongs to C$_3$.
 }
\label{acc}
\end{figure}

 We generated 6,000 spectra for each class, producing a data set containing 30,000 simulated spectra to train and test the model. From these, 60\% of the data set was used for training, and 40\% was used for testing. The optimal model was achieved for the values of the parameters displayed in Table~\ref{tab:table1}. The model's performance was evaluated based on the training/testing accuracy curves illustrated in the panel (a) of Fig.~\ref{acc}.

 The performance of the DNN classifier, as shown in panel (a) of Fig.~\ref{acc}, was assessed in terms of accuracy. The accuracy, as the metric used to evaluate the classification model, is the number of correct predictions out of the total number of predictions. An accuracy equal to 1 stands for the perfect performance of a model, and 0 stands for complete failure. As shown in the figure, the model achieves an accuracy between 0.7 and 0.75 after training for 40 epochs. Panel (b) in Fig.~\ref{acc} displays the confusion matrix, which gives information about the classifier's performance in assigning each simulated spectrum to the correct class. The elements on the diagonal represent a normalized number of ideally classified spectra, and the off-diagonal elements represent the misclassified spectra. The first and the second class show a high number of misclassified spectra because those two classes have similar peaks in the low energy regime ($<2.5$~MeV). Hence, it is harder for the network to distinguish them, especially when the data set fluctuates significantly.
Each element of the confusion matrix is estimated from representative test sets, counting the number of distributions assigned to each class and normalized by the number of spectra in that class.

After the DNN is trained, we use it to in classify the experimental spectrum. For example, if the assigned class is C$_1$, this means that the spectrum is perceived to have two states around the positions described before.  If it is C$_4$, the spectrum is recognized as having four peaks.
\begin{table*}
  
   \caption{The table shows the hyper-parameters used to design the DNN classification model and our assumptions to generate the training data set. The first part of the table displays parameters that made the DNN architecture (i.e., numbers of layers and neurons in each layer). The second part (middle) shows other hyper-parameters and also shows the positions of the peaks ($E_i$) used in Breit-Wigner distribution, Eq.~\eqref{Eq31} which was multiplied with TM to obtain the training set. The last part consists of classes, C$_1$, C$_2$, C$_3$, C$_4$, and C$_5$ created in such a way that each class has distributions with a number of peaks and/or features different from other classes.}
    \label{tab:table1}
    \centering
  \begin{adjustbox}{width=.75\textwidth}
    \small
    \label{tab:table1}
    \begin{tabular}{ccr}
        \hline
     \multirow{2}{*}{} \textbf{DNN architecture} \\
      \toprule 
      \textbf{Layers} & \textbf{Number of neurons} & \textbf{Activation function}\\
      \midrule 
      Input Layer & 50 & ReLU\\
      1$^{st}$ Hidden layer & 300& ReLU\\
      2$^{nd}$ Hidden layer& 500 & ReLU\\
      Output layer&5& Softmax\\
      \hline
     \multirow{2}{*}{} \textbf{Other hyper-parameters} & Peak location (MeV)& Peak width (MeV)\\
    \toprule 
      Optimizer (Adam) & $0.00\leq E_1\leq 0.30$ &$0.08\leq \Gamma_1\leq 0.50$\\
      Epoch number (200)& $1.10< E_2\leq 1.80$ &$0.50 < \Gamma_2\leq 1.10$\\
      Batch number (20)&$1.90< E_3\leq 2.70$& $1.10 < \Gamma_3\leq 1.30$\\
      Learning rate (0.004) &$2.70< E_4\leq 4.00$ &$1.30< \Gamma_4\leq 1.80$\\
      Learning rate (0.004) &$4.00< E_5\leq 6.00$ &$1.80< \Gamma_5\leq 2.10$\\
     \hline
     \multirow{2}{*}{} \textbf{prediction of DNN classifier on experimental data} \\
    \toprule 
     \textbf{Class} & \textbf{Description}\\
       label & Peak Position (resonances states) \\
      \midrule 
      1. C$_1$ &  $E_1$ and $E_2$ \\
      2. C$_2$ &  $E_1$, $E_2$, and $E_3$ \\
      3. C$_3$ & $E_1$, $E_2$, and  $E_5$ \\
      4. C$_4$ & $E_1$, $E_2$, $E_3$ and $E_5$ \\
      5. C$_5$ &  elsewhere   \\
     \hline
    \end{tabular}
    \end{adjustbox}
\end{table*}
However, we only have one measured decay energy spectrum from the experiment investigating the three-body decay of $^{26} \mathrm{O}$ into $^{24} \mathrm{O}$ and 2n. A fair prediction is expected, when enough samples are passed to the DNN model. For that reason, we carry out error resampling for the measured spectrum to obtain a dataset that one can use in the model prediction. With this process, we have generated 10,000 samples of distributions and estimated the classes to which each spectrum from the resampling belongs. We evaluated the number of distributions predicted to be in a given class as a fraction of the distributions in that class per the total number of distributions, see panel (c) of Fig.~\ref{acc}. With this, the number of resonance states most likely there in the measured decay energy spectrum corresponds to the class with the highest fractional value. As evident in Fig.~\ref{acc}(c), more than 75\% of the total distributions used in the prediction 
belong to class C$_3$, which suggests within our statistical framework the presence of three peaks in the measured spectrum. The locations of those peaks are approximately equal to the mean values of parameters $E_i$ of C$_3$  at 0.15, 1.50, and 5.00~MeV.  The mean values of half-widths with which the classifier sorts the three peaks, are, respectively, 0.29, 0.80, and $1.85 \, \text{MeV}$.  Finally, these peaks correspond to the resonance states of $^{26}$O reported in Refs.~\cite{kondo2016nucleus,caesar2013beyond}.
\section{Conclusions and Outlook}
We applied the deblurring method, successful in optics and employing the RL algorithm, to the restoration of the energy spectrum from the three-body decay of $^{26}\mathrm{O}$. 
As presented here, the algorithm requires only the measured distribution in energy and the TM, with elements only labeled by energy, to operate.  Two-dimensional distributions of photons are typically employed in optics and such and higher dimensions in nuclear applications can be envisioned.  The inversion implicit in the algorithm is largely stabilized by the positive-definite probabilistic nature of the measured and restored distributions and of the TM elements. When significant noise is present in the deblurred distribution, though, a short wavelength instability may develop in the restored distribution in the limit of many restoration iterations.  With the relative noise growing with energy, due to fewer counts there, we stabilize that instability with an energy-dependent regularization in the individual restoration steps.

Ahead of the data, we tested the method in the restoration of a simulated decay energy spectrum without and with significant noise, as was illustrated in Figs.~\ref{fig:my_label} and~\ref{fig:my_label1}. Then, we applied the method to the measured energy spectrum of the three-body decay of $^{26}\mathrm{O}$. Three peaks were observed in restored spectrum.  Two of those were found in the low energy region, at about 0.15 and 1.5 MeV, which may be tied to the previously identified (0$^+$) and (2$^+$) states of $^{26}\mathrm{O}$. The third peak is located between 4 and 6~MeV in the restored spectrum, and such a peak was previously reported in Ref.~\cite{caesar2013beyond}.

Moreover, we built a deep neural network classification model with the same purpose as the deblurring technique: to identify resonance states of $^{26}\mathrm{O}$ from the measured decay energy spectrum. The DNN model estimates presence of three peaks in the spectrum at approximate mean positions of 0.15~MeV, 1.50~MeV for the first and second peaks, and at about 5.00~MeV for the third. The half-widths of these three peaks have been found to be approximately 0.29~MeV, 0.80~MeV, and 1.85~MeV, respectively.
The agreement between the two methods used in our analyses suggests that there may be three resonance states of $^{26}\mathrm{O}$ impacting the measured decay energy spectrum. In addition, the result from chi-square minimization, shown in Fig.~\ref{chisqrt}, suggests that three peaks suffice to describe the data.

Possible nuclear applications of the deblurring method we described, besides recovering decay energy spectra, and the aforementioned 3D distributions in heavy-ion collisions, can include restoration of emitting source distribution from particle correlations in heavy-ion collisions. The emitting source function gives information about spatial geometry and time development of the final stages of reactions \cite{bauer1993particle,heinz1996lifetimes,koonin1977proton,pratt1984pion}, as well as their phase-space \cite{bertsch_meson_1996} and thermodynamic characteristics \cite{chapman1995model}. 

This work was supported by the U.S.\ Department of Energy Office of Science under Grant No.\ DE-SC0019209, the U.S.\ National Science Foundation Grant No.\ PHY-2012040, and the U.S.\ National Science Foundation CSSI Program Grant No.\ OAC-2004601 (BAND Collaboration).
\newpage
\bibliographystyle{apsrev}
\bibliography{dnn}

\begin{thebibliography}{39}
\expandafter\ifx\csname natexlab\endcsname\relax\def\natexlab#1{#1}\fi
\expandafter\ifx\csname bibnamefont\endcsname\relax
  \def\bibnamefont#1{#1}\fi
\expandafter\ifx\csname bibfnamefont\endcsname\relax
  \def\bibfnamefont#1{#1}\fi
\expandafter\ifx\csname citenamefont\endcsname\relax
  \def\citenamefont#1{#1}\fi
\expandafter\ifx\csname url\endcsname\relax
  \def\url#1{\texttt{#1}}\fi
\expandafter\ifx\csname urlprefix\endcsname\relax\def\urlprefix{URL }\fi
\providecommand{\bibinfo}[2]{#2}
\providecommand{\eprint}[2][]{\url{#2}}

\bibitem[{\citenamefont{Redpath}(2019)}]{redpath2019measuring}
\bibinfo{author}{\bibfnamefont{T.}~\bibnamefont{Redpath}},
  \emph{\bibinfo{title}{Measuring the Half-life of O-26}}
  (\bibinfo{publisher}{Michigan State University. Physics},
  \bibinfo{year}{2019}).

\bibitem[{\citenamefont{Caesar et~al.}(2013)\citenamefont{Caesar, Simonis,
  Adachi, Aksyutina, Alcantara, Altstadt, Alvarez-Pol, Ashwood, Aumann,
  Avdeichikov et~al.}}]{caesar2013beyond}
\bibinfo{author}{\bibfnamefont{C.}~\bibnamefont{Caesar}},
  \bibinfo{author}{\bibfnamefont{J.}~\bibnamefont{Simonis}},
  \bibinfo{author}{\bibfnamefont{T.}~\bibnamefont{Adachi}},
  \bibinfo{author}{\bibfnamefont{Y.}~\bibnamefont{Aksyutina}},
  \bibinfo{author}{\bibfnamefont{J.}~\bibnamefont{Alcantara}},
  \bibinfo{author}{\bibfnamefont{S.}~\bibnamefont{Altstadt}},
  \bibinfo{author}{\bibfnamefont{H.}~\bibnamefont{Alvarez-Pol}},
  \bibinfo{author}{\bibfnamefont{N.}~\bibnamefont{Ashwood}},
  \bibinfo{author}{\bibfnamefont{T.}~\bibnamefont{Aumann}},
  \bibinfo{author}{\bibfnamefont{V.}~\bibnamefont{Avdeichikov}},
  \bibnamefont{et~al.}, \bibinfo{journal}{Phys. Rev. C}
  \textbf{\bibinfo{volume}{88}}, \bibinfo{pages}{034313}
  (\bibinfo{year}{2013}).

\bibitem[{\citenamefont{Chrisman et~al.}(2021)\citenamefont{Chrisman, Kuchera,
  Baumann, Blake, Brown, Brown, Cochran, DeYoung, Finck, Frank
  et~al.}}]{chrisman2021neutron}
\bibinfo{author}{\bibfnamefont{D.}~\bibnamefont{Chrisman}},
  \bibinfo{author}{\bibfnamefont{A.~N.} \bibnamefont{Kuchera}},
  \bibinfo{author}{\bibfnamefont{T.}~\bibnamefont{Baumann}},
  \bibinfo{author}{\bibfnamefont{A.}~\bibnamefont{Blake}},
  \bibinfo{author}{\bibfnamefont{B.~A.} \bibnamefont{Brown}},
  \bibinfo{author}{\bibfnamefont{J.}~\bibnamefont{Brown}},
  \bibinfo{author}{\bibfnamefont{C.}~\bibnamefont{Cochran}},
  \bibinfo{author}{\bibfnamefont{P.~A.} \bibnamefont{DeYoung}},
  \bibinfo{author}{\bibfnamefont{J.~E.} \bibnamefont{Finck}},
  \bibinfo{author}{\bibfnamefont{N.}~\bibnamefont{Frank}},
  \bibnamefont{et~al.}, \bibinfo{journal}{Phys. Rev. C}
  \textbf{\bibinfo{volume}{104}}, \bibinfo{pages}{034313}
  (\bibinfo{year}{2021}),
  \urlprefix\url{https://link.aps.org/doi/10.1103/PhysRevC.104.034313}.

\bibitem[{\citenamefont{Revel et~al.}(2020)\citenamefont{Revel, Sorlin,
  Marqu\'es, Kondo, Kahlbow, Nakamura, Orr, Nowacki, Tostevin, Yuan
  et~al.}}]{revel2020f28}
\bibinfo{author}{\bibfnamefont{A.}~\bibnamefont{Revel}},
  \bibinfo{author}{\bibfnamefont{O.}~\bibnamefont{Sorlin}},
  \bibinfo{author}{\bibfnamefont{F.~M.} \bibnamefont{Marqu\'es}},
  \bibinfo{author}{\bibfnamefont{Y.}~\bibnamefont{Kondo}},
  \bibinfo{author}{\bibfnamefont{J.}~\bibnamefont{Kahlbow}},
  \bibinfo{author}{\bibfnamefont{T.}~\bibnamefont{Nakamura}},
  \bibinfo{author}{\bibfnamefont{N.~A.} \bibnamefont{Orr}},
  \bibinfo{author}{\bibfnamefont{F.}~\bibnamefont{Nowacki}},
  \bibinfo{author}{\bibfnamefont{J.~A.} \bibnamefont{Tostevin}},
  \bibinfo{author}{\bibfnamefont{C.~X.} \bibnamefont{Yuan}},
  \bibnamefont{et~al.} (\bibinfo{collaboration}{SAMURAI21 collaboration}),
  \bibinfo{journal}{Phys. Rev. Lett.} \textbf{\bibinfo{volume}{124}},
  \bibinfo{pages}{152502} (\bibinfo{year}{2020}),
  \urlprefix\url{https://link.aps.org/doi/10.1103/PhysRevLett.124.152502}.

\bibitem[{\citenamefont{Leblond et~al.}(2018)\citenamefont{Leblond, Marqu\'es,
  Gibelin, Orr, Kondo, Nakamura, Bonnard, Michel, Achouri, Aumann
  et~al.}}]{leblond2018b21}
\bibinfo{author}{\bibfnamefont{S.}~\bibnamefont{Leblond}},
  \bibinfo{author}{\bibfnamefont{F.~M.} \bibnamefont{Marqu\'es}},
  \bibinfo{author}{\bibfnamefont{J.}~\bibnamefont{Gibelin}},
  \bibinfo{author}{\bibfnamefont{N.~A.} \bibnamefont{Orr}},
  \bibinfo{author}{\bibfnamefont{Y.}~\bibnamefont{Kondo}},
  \bibinfo{author}{\bibfnamefont{T.}~\bibnamefont{Nakamura}},
  \bibinfo{author}{\bibfnamefont{J.}~\bibnamefont{Bonnard}},
  \bibinfo{author}{\bibfnamefont{N.}~\bibnamefont{Michel}},
  \bibinfo{author}{\bibfnamefont{N.~L.} \bibnamefont{Achouri}},
  \bibinfo{author}{\bibfnamefont{T.}~\bibnamefont{Aumann}},
  \bibnamefont{et~al.}, \bibinfo{journal}{Phys. Rev. Lett.}
  \textbf{\bibinfo{volume}{121}}, \bibinfo{pages}{262502}
  (\bibinfo{year}{2018}),
  \urlprefix\url{https://link.aps.org/doi/10.1103/PhysRevLett.121.262502}.

\bibitem[{\citenamefont{Lane and Thomas}(1958)}]{rmatrix}
\bibinfo{author}{\bibfnamefont{A.~M.} \bibnamefont{Lane}} \bibnamefont{and}
  \bibinfo{author}{\bibfnamefont{R.~G.} \bibnamefont{Thomas}},
  \bibinfo{journal}{Rev. Mod. Phys.} \textbf{\bibinfo{volume}{30}},
  \bibinfo{pages}{257} (\bibinfo{year}{1958}),
  \urlprefix\url{https://link.aps.org/doi/10.1103/RevModPhys.30.257}.

\bibitem[{\citenamefont{Richardson}(1972)}]{richardson1972bayesian}
\bibinfo{author}{\bibfnamefont{W.~H.} \bibnamefont{Richardson}},
  \bibinfo{journal}{JoSA} \textbf{\bibinfo{volume}{62}}, \bibinfo{pages}{55}
  (\bibinfo{year}{1972}).

\bibitem[{\citenamefont{Lucy}(1974)}]{lucy1974iterative}
\bibinfo{author}{\bibfnamefont{L.~B.} \bibnamefont{Lucy}},
  \bibinfo{journal}{Astron. J.} \textbf{\bibinfo{volume}{79}},
  \bibinfo{pages}{745} (\bibinfo{year}{1974}).

\bibitem[{\citenamefont{Thi{\'e}baut et~al.}(2016)\citenamefont{Thi{\'e}baut,
  D{\'e}nis, Soulez, and Mourya}}]{thiebaut2016spatially}
\bibinfo{author}{\bibfnamefont{{\'E}.}~\bibnamefont{Thi{\'e}baut}},
  \bibinfo{author}{\bibfnamefont{L.}~\bibnamefont{D{\'e}nis}},
  \bibinfo{author}{\bibfnamefont{F.}~\bibnamefont{Soulez}}, \bibnamefont{and}
  \bibinfo{author}{\bibfnamefont{R.}~\bibnamefont{Mourya}}, in
  \emph{\bibinfo{booktitle}{Adaptive Optics Systems V}}
  (\bibinfo{organization}{International Society for Optics and Photonics},
  \bibinfo{year}{2016}), vol. \bibinfo{volume}{9909}, p.
  \bibinfo{pages}{99097N}.

\bibitem[{\citenamefont{Al-Ameen and Sulong}(2015)}]{al2015deblurring}
\bibinfo{author}{\bibfnamefont{Z.}~\bibnamefont{Al-Ameen}} \bibnamefont{and}
  \bibinfo{author}{\bibfnamefont{G.}~\bibnamefont{Sulong}},
  \bibinfo{journal}{Interdiscipl. Sci.: Comput. Life Sci.,}
  \textbf{\bibinfo{volume}{7}}, \bibinfo{pages}{319} (\bibinfo{year}{2015}).

\bibitem[{\citenamefont{D'Agostini}(1995)}]{dagostini_multidimensional_1995}
\bibinfo{author}{\bibfnamefont{G.}~\bibnamefont{D'Agostini}},
  \bibinfo{journal}{Nucl. Instrum. Methods. Phys. Res. A}
  \textbf{\bibinfo{volume}{362}}, \bibinfo{pages}{487} (\bibinfo{year}{1995}),
  ISSN \bibinfo{issn}{0168-9002},
  \urlprefix\url{https://www.sciencedirect.com/science/article/pii/016890029500274X}.

\bibitem[{\citenamefont{Danielewicz and
  Kurata-Nishimura}(2022)}]{danielewicz2022deblurring}
\bibinfo{author}{\bibfnamefont{P.}~\bibnamefont{Danielewicz}} \bibnamefont{and}
  \bibinfo{author}{\bibfnamefont{M.}~\bibnamefont{Kurata-Nishimura}},
  \bibinfo{journal}{Phys. Rev. C} \textbf{\bibinfo{volume}{105}},
  \bibinfo{pages}{034608} (\bibinfo{year}{2022}).

\bibitem[{\citenamefont{Grech et~al.}(2008)\citenamefont{Grech, Cassar, Muscat,
  Camilleri, Fabri, Zervakis, Xanthopoulos, Sakkalis, and
  Vanrumste}}]{grech2008review}
\bibinfo{author}{\bibfnamefont{R.}~\bibnamefont{Grech}},
  \bibinfo{author}{\bibfnamefont{T.}~\bibnamefont{Cassar}},
  \bibinfo{author}{\bibfnamefont{J.}~\bibnamefont{Muscat}},
  \bibinfo{author}{\bibfnamefont{K.~P.} \bibnamefont{Camilleri}},
  \bibinfo{author}{\bibfnamefont{S.~G.} \bibnamefont{Fabri}},
  \bibinfo{author}{\bibfnamefont{M.}~\bibnamefont{Zervakis}},
  \bibinfo{author}{\bibfnamefont{P.}~\bibnamefont{Xanthopoulos}},
  \bibinfo{author}{\bibfnamefont{V.}~\bibnamefont{Sakkalis}}, \bibnamefont{and}
  \bibinfo{author}{\bibfnamefont{B.}~\bibnamefont{Vanrumste}},
  \bibinfo{journal}{J. Neuroengineering.Rehabil.} \textbf{\bibinfo{volume}{5}},
  \bibinfo{pages}{1} (\bibinfo{year}{2008}),
  \urlprefix\url{https://doi.org/10.1186/1743-0003-5-25}.

\bibitem[{\citenamefont{Fister et~al.}(2007)\citenamefont{Fister, Seidler,
  Rehr, Kas, Elam, Cross, and Nagle}}]{fister2007deconvolving}
\bibinfo{author}{\bibfnamefont{T.~T.} \bibnamefont{Fister}},
  \bibinfo{author}{\bibfnamefont{G.~T.} \bibnamefont{Seidler}},
  \bibinfo{author}{\bibfnamefont{J.~J.} \bibnamefont{Rehr}},
  \bibinfo{author}{\bibfnamefont{J.~J.} \bibnamefont{Kas}},
  \bibinfo{author}{\bibfnamefont{W.~T.} \bibnamefont{Elam}},
  \bibinfo{author}{\bibfnamefont{J.~O.} \bibnamefont{Cross}}, \bibnamefont{and}
  \bibinfo{author}{\bibfnamefont{K.~P.} \bibnamefont{Nagle}},
  \bibinfo{journal}{Phys. Rev. B} \textbf{\bibinfo{volume}{75}},
  \bibinfo{pages}{174106} (\bibinfo{year}{2007}),
  \urlprefix\url{https://link.aps.org/doi/10.1103/PhysRevB.75.174106}.

\bibitem[{\citenamefont{Dey et~al.}(2006)\citenamefont{Dey, Blanc-Feraud,
  Zimmer, Roux, Kam, Olivo-Marin, and Zerubia}}]{dey2006richardson}
\bibinfo{author}{\bibfnamefont{N.}~\bibnamefont{Dey}},
  \bibinfo{author}{\bibfnamefont{L.}~\bibnamefont{Blanc-Feraud}},
  \bibinfo{author}{\bibfnamefont{C.}~\bibnamefont{Zimmer}},
  \bibinfo{author}{\bibfnamefont{P.}~\bibnamefont{Roux}},
  \bibinfo{author}{\bibfnamefont{Z.}~\bibnamefont{Kam}},
  \bibinfo{author}{\bibfnamefont{J.-C.} \bibnamefont{Olivo-Marin}},
  \bibnamefont{and} \bibinfo{author}{\bibfnamefont{J.}~\bibnamefont{Zerubia}},
  \bibinfo{journal}{Microsc. Res. Tech.} \textbf{\bibinfo{volume}{69}},
  \bibinfo{pages}{260} (\bibinfo{year}{2006}).

\bibitem[{\citenamefont{Vargas et~al.}(2013)\citenamefont{Vargas, Benlliure,
  and Caamaño}}]{vargas_unfolding_2013}
\bibinfo{author}{\bibfnamefont{J.}~\bibnamefont{Vargas}},
  \bibinfo{author}{\bibfnamefont{J.}~\bibnamefont{Benlliure}},
  \bibnamefont{and} \bibinfo{author}{\bibfnamefont{M.}~\bibnamefont{Caamaño}},
  \bibinfo{journal}{Nucl. Instrum. Methodes. Phys. Res. Sect., A:}
  \textbf{\bibinfo{volume}{707}}, \bibinfo{pages}{16} (\bibinfo{year}{2013}),
  \urlprefix\url{https://www.sciencedirect.com/science/article/pii/S016890021201635X}.

\bibitem[{\citenamefont{Rudin et~al.}(1992)\citenamefont{Rudin, Osher, and
  Fatemi}}]{rudin1992nonlinear}
\bibinfo{author}{\bibfnamefont{L.~I.} \bibnamefont{Rudin}},
  \bibinfo{author}{\bibfnamefont{S.}~\bibnamefont{Osher}}, \bibnamefont{and}
  \bibinfo{author}{\bibfnamefont{E.}~\bibnamefont{Fatemi}},
  \bibinfo{journal}{Physica D} \textbf{\bibinfo{volume}{60}},
  \bibinfo{pages}{259} (\bibinfo{year}{1992}).

\bibitem[{\citenamefont{Lawrence et~al.}(1997)\citenamefont{Lawrence, Giles,
  Tsoi, and Back}}]{554195}
\bibinfo{author}{\bibfnamefont{S.}~\bibnamefont{Lawrence}},
  \bibinfo{author}{\bibfnamefont{C.}~\bibnamefont{Giles}},
  \bibinfo{author}{\bibfnamefont{A.~C.} \bibnamefont{Tsoi}}, \bibnamefont{and}
  \bibinfo{author}{\bibfnamefont{A.}~\bibnamefont{Back}},
  \bibinfo{journal}{IEEE Trans. Neural Networks} \textbf{\bibinfo{volume}{8}},
  \bibinfo{pages}{98} (\bibinfo{year}{1997}).

\bibitem[{\citenamefont{Nassif et~al.}(2019)\citenamefont{Nassif, Shahin,
  Attili, Azzeh, and Shaalan}}]{8632885}
\bibinfo{author}{\bibfnamefont{A.~B.} \bibnamefont{Nassif}},
  \bibinfo{author}{\bibfnamefont{I.}~\bibnamefont{Shahin}},
  \bibinfo{author}{\bibfnamefont{I.}~\bibnamefont{Attili}},
  \bibinfo{author}{\bibfnamefont{M.}~\bibnamefont{Azzeh}}, \bibnamefont{and}
  \bibinfo{author}{\bibfnamefont{K.}~\bibnamefont{Shaalan}},
  \bibinfo{journal}{IEEE Access} \textbf{\bibinfo{volume}{7}},
  \bibinfo{pages}{19143} (\bibinfo{year}{2019}).

\bibitem[{\citenamefont{Guest et~al.}(2018)\citenamefont{Guest, Cranmer, and
  Whiteson}}]{guest2018deep}
\bibinfo{author}{\bibfnamefont{D.}~\bibnamefont{Guest}},
  \bibinfo{author}{\bibfnamefont{K.}~\bibnamefont{Cranmer}}, \bibnamefont{and}
  \bibinfo{author}{\bibfnamefont{D.}~\bibnamefont{Whiteson}},
  \bibinfo{journal}{Ann. Rev. Nucl. Part. Sci.,} \textbf{\bibinfo{volume}{68}},
  \bibinfo{pages}{161} (\bibinfo{year}{2018}).

\bibitem[{\citenamefont{Matchev and Shyamsundar}(2021)}]{matchev2021thickbrick}
\bibinfo{author}{\bibfnamefont{K.~T.} \bibnamefont{Matchev}} \bibnamefont{and}
  \bibinfo{author}{\bibfnamefont{P.}~\bibnamefont{Shyamsundar}},
  \bibinfo{journal}{J. High Energ. Phys.} \textbf{\bibinfo{volume}{2021}},
  \bibinfo{pages}{1} (\bibinfo{year}{2021}),
  \urlprefix\url{https://doi.org/10.1007/JHEP03(2021)291}.

\bibitem[{\citenamefont{Carleo and Troyer}(2017)}]{carleo2017solving}
\bibinfo{author}{\bibfnamefont{G.}~\bibnamefont{Carleo}} \bibnamefont{and}
  \bibinfo{author}{\bibfnamefont{M.}~\bibnamefont{Troyer}},
  \bibinfo{journal}{Science} \textbf{\bibinfo{volume}{355}},
  \bibinfo{pages}{602} (\bibinfo{year}{2017}).

\bibitem[{\citenamefont{Whiteson and Whiteson}(2009)}]{whiteson2009machine}
\bibinfo{author}{\bibfnamefont{S.}~\bibnamefont{Whiteson}} \bibnamefont{and}
  \bibinfo{author}{\bibfnamefont{D.}~\bibnamefont{Whiteson}},
  \bibinfo{journal}{Eng. Appl. Artif. Intell.,} \textbf{\bibinfo{volume}{22}},
  \bibinfo{pages}{1203} (\bibinfo{year}{2009}),
  \urlprefix\url{https://www.sciencedirect.com/science/article/pii/S0952197609000827}.

\bibitem[{\citenamefont{Fujimoto et~al.}(2020)\citenamefont{Fujimoto,
  Fukushima, and Murase}}]{fujimoto2020mapping}
\bibinfo{author}{\bibfnamefont{Y.}~\bibnamefont{Fujimoto}},
  \bibinfo{author}{\bibfnamefont{K.}~\bibnamefont{Fukushima}},
  \bibnamefont{and} \bibinfo{author}{\bibfnamefont{K.}~\bibnamefont{Murase}},
  \bibinfo{journal}{Phys. Rev. D} \textbf{\bibinfo{volume}{101}},
  \bibinfo{pages}{054016} (\bibinfo{year}{2020}).

\bibitem[{\citenamefont{Bedaque et~al.}(2021)\citenamefont{Bedaque, Boehnlein,
  Cromaz, Diefenthaler, Elouadrhiri, Horn, Kuchera, Lawrence, Lee, Lidia
  et~al.}}]{bedaque2021ai}
\bibinfo{author}{\bibfnamefont{P.}~\bibnamefont{Bedaque}},
  \bibinfo{author}{\bibfnamefont{A.}~\bibnamefont{Boehnlein}},
  \bibinfo{author}{\bibfnamefont{M.}~\bibnamefont{Cromaz}},
  \bibinfo{author}{\bibfnamefont{M.}~\bibnamefont{Diefenthaler}},
  \bibinfo{author}{\bibfnamefont{L.}~\bibnamefont{Elouadrhiri}},
  \bibinfo{author}{\bibfnamefont{T.}~\bibnamefont{Horn}},
  \bibinfo{author}{\bibfnamefont{M.}~\bibnamefont{Kuchera}},
  \bibinfo{author}{\bibfnamefont{D.}~\bibnamefont{Lawrence}},
  \bibinfo{author}{\bibfnamefont{D.}~\bibnamefont{Lee}},
  \bibinfo{author}{\bibfnamefont{S.}~\bibnamefont{Lidia}},
  \bibnamefont{et~al.}, \bibinfo{journal}{Eur. Phys. J. A}
  \textbf{\bibinfo{volume}{57}}, \bibinfo{pages}{1} (\bibinfo{year}{2021}).

\bibitem[{\citenamefont{Redpath et~al.}(2020)\citenamefont{Redpath, Baumann,
  Brown, Chrisman, DeYoung, Frank, Gu{\`e}ye, Kuchera, Liu, Persch
  et~al.}}]{redpath2020new}
\bibinfo{author}{\bibfnamefont{T.}~\bibnamefont{Redpath}},
  \bibinfo{author}{\bibfnamefont{T.}~\bibnamefont{Baumann}},
  \bibinfo{author}{\bibfnamefont{J.}~\bibnamefont{Brown}},
  \bibinfo{author}{\bibfnamefont{D.}~\bibnamefont{Chrisman}},
  \bibinfo{author}{\bibfnamefont{P.}~\bibnamefont{DeYoung}},
  \bibinfo{author}{\bibfnamefont{N.}~\bibnamefont{Frank}},
  \bibinfo{author}{\bibfnamefont{P.}~\bibnamefont{Gu{\`e}ye}},
  \bibinfo{author}{\bibfnamefont{A.}~\bibnamefont{Kuchera}},
  \bibinfo{author}{\bibfnamefont{H.}~\bibnamefont{Liu}},
  \bibinfo{author}{\bibfnamefont{C.}~\bibnamefont{Persch}},
  \bibnamefont{et~al.}, \bibinfo{journal}{Nucl. Instrum. Methods. Phys. Res.
  Sec., A} \textbf{\bibinfo{volume}{977}}, \bibinfo{pages}{164284}
  (\bibinfo{year}{2020}).

\bibitem[{\citenamefont{Bohm and Zech}(2010)}]{bohm_introduction_2010}
\bibinfo{author}{\bibfnamefont{G.}~\bibnamefont{Bohm}} \bibnamefont{and}
  \bibinfo{author}{\bibfnamefont{G.}~\bibnamefont{Zech}},
  \emph{\bibinfo{title}{Introduction to statistics and data analysis for
  physicists}} (\bibinfo{publisher}{DESY}, \bibinfo{year}{2010}), ISBN
  \bibinfo{isbn}{978-3-935702-41-6}.

\bibitem[{\citenamefont{Kohley et~al.}(2013)\citenamefont{Kohley, Baumann,
  Bazin, Christian, DeYoung, Finck, Frank, Jones, Lunderberg, Luther
  et~al.}}]{kohley2013study}
\bibinfo{author}{\bibfnamefont{Z.}~\bibnamefont{Kohley}},
  \bibinfo{author}{\bibfnamefont{T.}~\bibnamefont{Baumann}},
  \bibinfo{author}{\bibfnamefont{D.}~\bibnamefont{Bazin}},
  \bibinfo{author}{\bibfnamefont{G.}~\bibnamefont{Christian}},
  \bibinfo{author}{\bibfnamefont{P.~A.} \bibnamefont{DeYoung}},
  \bibinfo{author}{\bibfnamefont{J.~E.} \bibnamefont{Finck}},
  \bibinfo{author}{\bibfnamefont{N.}~\bibnamefont{Frank}},
  \bibinfo{author}{\bibfnamefont{M.}~\bibnamefont{Jones}},
  \bibinfo{author}{\bibfnamefont{E.}~\bibnamefont{Lunderberg}},
  \bibinfo{author}{\bibfnamefont{B.}~\bibnamefont{Luther}},
  \bibnamefont{et~al.}, \bibinfo{journal}{Phys. Rev. Lett.}
  \textbf{\bibinfo{volume}{110}}, \bibinfo{pages}{152501}
  (\bibinfo{year}{2013}),
  \urlprefix\url{https://link.aps.org/doi/10.1103/PhysRevLett.110.152501}.

\bibitem[{\citenamefont{Kondo et~al.}(2016)\citenamefont{Kondo, Nakamura,
  Tanaka, Minakata, Ogoshi, Orr, Achouri, Aumann, Baba, Delaunay
  et~al.}}]{kondo2016nucleus}
\bibinfo{author}{\bibfnamefont{Y.}~\bibnamefont{Kondo}},
  \bibinfo{author}{\bibfnamefont{T.}~\bibnamefont{Nakamura}},
  \bibinfo{author}{\bibfnamefont{R.}~\bibnamefont{Tanaka}},
  \bibinfo{author}{\bibfnamefont{R.}~\bibnamefont{Minakata}},
  \bibinfo{author}{\bibfnamefont{S.}~\bibnamefont{Ogoshi}},
  \bibinfo{author}{\bibfnamefont{N.}~\bibnamefont{Orr}},
  \bibinfo{author}{\bibfnamefont{N.}~\bibnamefont{Achouri}},
  \bibinfo{author}{\bibfnamefont{T.}~\bibnamefont{Aumann}},
  \bibinfo{author}{\bibfnamefont{H.}~\bibnamefont{Baba}},
  \bibinfo{author}{\bibfnamefont{F.}~\bibnamefont{Delaunay}},
  \bibnamefont{et~al.}, \bibinfo{journal}{Phys. Rev. Lett.}
  \textbf{\bibinfo{volume}{116}}, \bibinfo{pages}{102503}
  (\bibinfo{year}{2016}).

\bibitem[{\citenamefont{Daubechies et~al.}(2019)\citenamefont{Daubechies,
  DeVore, Foucart, Hanin, and Petrova}}]{daubechies2019nonlinear}
\bibinfo{author}{\bibfnamefont{I.}~\bibnamefont{Daubechies}},
  \bibinfo{author}{\bibfnamefont{R.}~\bibnamefont{DeVore}},
  \bibinfo{author}{\bibfnamefont{S.}~\bibnamefont{Foucart}},
  \bibinfo{author}{\bibfnamefont{B.}~\bibnamefont{Hanin}}, \bibnamefont{and}
  \bibinfo{author}{\bibfnamefont{G.}~\bibnamefont{Petrova}},
  \bibinfo{journal}{arXiv preprint arXiv:1905.02199}  (\bibinfo{year}{2019}).

\bibitem[{\citenamefont{Sharma et~al.}(2017)\citenamefont{Sharma, Sharma, and
  Athaiya}}]{sharma2017activation}
\bibinfo{author}{\bibfnamefont{S.}~\bibnamefont{Sharma}},
  \bibinfo{author}{\bibfnamefont{S.}~\bibnamefont{Sharma}}, \bibnamefont{and}
  \bibinfo{author}{\bibfnamefont{A.}~\bibnamefont{Athaiya}},
  \bibinfo{journal}{Towards Data Science} \textbf{\bibinfo{volume}{6}},
  \bibinfo{pages}{310} (\bibinfo{year}{2017}),
  \urlprefix\url{https://towardsdatascience.com/activation-functions-neural-networks-1cbd9f8d91d6}.

\bibitem[{\citenamefont{Rusiecki}(2019)}]{rusiecki2019trimmed}
\bibinfo{author}{\bibfnamefont{A.}~\bibnamefont{Rusiecki}},
  \bibinfo{journal}{Electron. Lett.} \textbf{\bibinfo{volume}{55}},
  \bibinfo{pages}{319} (\bibinfo{year}{2019}),
  \urlprefix\url{https://doi.org/10.1049/el.2018.7980}.

\bibitem[{\citenamefont{Kingma and Ba}(2014)}]{kingma2014adam}
\bibinfo{author}{\bibfnamefont{D.~P.} \bibnamefont{Kingma}} \bibnamefont{and}
  \bibinfo{author}{\bibfnamefont{J.}~\bibnamefont{Ba}}, \bibinfo{journal}{arXiv
  preprint arXiv:1412.6980}  (\bibinfo{year}{2014}).

\bibitem[{\citenamefont{Bauer}(1993)}]{bauer1993particle}
\bibinfo{author}{\bibfnamefont{W.}~\bibnamefont{Bauer}},
  \bibinfo{journal}{Prog. Part. Nucl. Phys.} \textbf{\bibinfo{volume}{30}},
  \bibinfo{pages}{45} (\bibinfo{year}{1993}), ISSN \bibinfo{issn}{0146-6410},
  \urlprefix\url{https://www.sciencedirect.com/science/article/pii/014664109390005Z}.

\bibitem[{\citenamefont{Heinz et~al.}(1996)\citenamefont{Heinz,
  Tom{\'a}{\v{s}}ik, Wiedemann, and Wu}}]{heinz1996lifetimes}
\bibinfo{author}{\bibfnamefont{U.}~\bibnamefont{Heinz}},
  \bibinfo{author}{\bibfnamefont{B.}~\bibnamefont{Tom{\'a}{\v{s}}ik}},
  \bibinfo{author}{\bibfnamefont{U.}~\bibnamefont{Wiedemann}},
  \bibnamefont{and} \bibinfo{author}{\bibfnamefont{Y.-F.} \bibnamefont{Wu}},
  \bibinfo{journal}{Phys. Lett. B} \textbf{\bibinfo{volume}{382}},
  \bibinfo{pages}{181} (\bibinfo{year}{1996}).

\bibitem[{\citenamefont{Koonin}(1977)}]{koonin1977proton}
\bibinfo{author}{\bibfnamefont{S.~E.} \bibnamefont{Koonin}},
  \bibinfo{journal}{Phys. Lett. B} \textbf{\bibinfo{volume}{70}},
  \bibinfo{pages}{43} (\bibinfo{year}{1977}).

\bibitem[{\citenamefont{Pratt}(1984)}]{pratt1984pion}
\bibinfo{author}{\bibfnamefont{S.}~\bibnamefont{Pratt}},
  \bibinfo{journal}{Phys. Rev. Lett.} \textbf{\bibinfo{volume}{53}},
  \bibinfo{pages}{1219} (\bibinfo{year}{1984}).

\bibitem[{\citenamefont{Bertsch}(1996)}]{bertsch_meson_1996}
\bibinfo{author}{\bibfnamefont{G.~F.} \bibnamefont{Bertsch}},
  \bibinfo{journal}{Phys. Rev. Lett.} \textbf{\bibinfo{volume}{77}},
  \bibinfo{pages}{789} (\bibinfo{year}{1996}),
  \urlprefix\url{https://link.aps.org/doi/10.1103/PhysRevLett.77.789}.

\bibitem[{\citenamefont{Chapman et~al.}(1995)\citenamefont{Chapman, Scotto, and
  Heinz}}]{chapman1995model}
\bibinfo{author}{\bibfnamefont{S.}~\bibnamefont{Chapman}},
  \bibinfo{author}{\bibfnamefont{P.}~\bibnamefont{Scotto}}, \bibnamefont{and}
  \bibinfo{author}{\bibfnamefont{U.}~\bibnamefont{Heinz}},
  \bibinfo{journal}{APH.N.S. Heavy Ion Physics} \textbf{\bibinfo{volume}{1}},
  \bibinfo{pages}{1} (\bibinfo{year}{1995}),
  \urlprefix\url{https://doi.org/10.1007/BF03053639}.

\end{thebibliography}
\end{document}